\documentclass[11pt,a4paper]{article}
\pdfoutput=1
\usepackage{jheppub}

\usepackage[numbers,sort&compress]{natbib}
\usepackage{subcaption}
\usepackage{enumerate}
\usepackage{xifthen} % if-then-else statements
\usepackage{tikz}
\usepackage{dsfont}
\usepackage{mathtools}

\usepackage{sectsty}
\allsectionsfont{\boldmath}

\renewcommand*{\a}{\alpha}

\newcommand*{\N}{\mathcal{N}}

\newcommand*{\beq}{\begin{equation}}
\newcommand*{\eeq}{\end{equation}}
\newcommand*{\bea}{\begin{eqnarray}}
\newcommand*{\eea}{\end{eqnarray}}

\newcommand\mf[1]{{\mathfrak{#1}}}

\definecolor{cardinal}{rgb}{0.6,0,0}
\definecolor{darkgreen}{rgb}{0,0.5,0}
\definecolor{golden}{rgb}{0.92, 0.7, 0}
\definecolor{midnight}{rgb}{0, 0, 0.5}
\definecolor{darkblue}{rgb}{0.2, 0, 0.8}

\newcommand*{\C}[2][]{
\ifthenelse{\isempty{#1}}{
{C_{#2}}
}{
	{C_{#2,\,#1}}
}
}
\newcommand*{\CF}[2][]{
\ifthenelse{\isempty{#1}}{
{F_{#2}}
}{
	{F_{#2,\,#1}}
}
}

\newcommand\eq[1]{eq.~(\ref{eq:#1})}
\newcommand\tbl[1]{table~\ref{tab:#1}}
\newcommand{\sn}[1]{section~\ref{sec:#1}}
\newcommand{\fig}[1]{figure~\ref{fig:#1}}
\newcommand{\app}[1]{appendix~\ref{app:#1}}

\title{\boldmath From Large to Small $\N=(4,4)$ Superconformal Surface Defects in Holographic 6d SCFTs}
\author[1]{Pietro Capuozzo}
\author[2]{\!\!, John Estes}
\author[3]{\!\!, Brandon Robinson}
\author[1]{\!\!,  and Benjamin Suzzoni}
\affiliation[1]{STAG Research Centre, Physics and Astronomy, University of Southampton,\\
    Highfield, Southampton SO17 1BJ, United Kingdom}
    \affiliation[2]{Department of Chemistry and Physics, SUNY Old Westbury, Old Westbury, NY, United States}
    \affiliation[3]{INFN, sezione di Milano-Bicocca, Piazza della Scienza 3, I-20126 Milano, Italy}
    \emailAdd{p.capuozzo@soton.ac.uk}
    \emailAdd{estesj@oldwestbury.edu}
\emailAdd{brandon.robinson@mib.infn.it}
\emailAdd{b.suzzoni@soton.ac.uk}
\preprint{}
\abstract{Two-dimensional (2d) $\N=(4,4)$ Lie superalgebras can be either ``small'' or ``large'', meaning their R-symmetry is either $\mathfrak{so}(4)$ or $\mathfrak{so}(4) \oplus \mathfrak{so}(4)$, respectively. Both cases admit a superconformal extension and fit into the one-parameter family $\mathfrak{d}\left(2,1;\gamma\right)\oplus \mathfrak{d}\left(2,1;\gamma\right)$, with parameter $\gamma \in (-\infty,\infty)$.  The large algebra corresponds to generic values of $\gamma$, while the small case corresponds to a degeneration limit with $\gamma \to -\infty$. In 11d supergravity, we study known solutions with superisometry algebra $\mathfrak{d}\left(2,1;\gamma\right)\oplus \mathfrak{d}\left(2,1;\gamma\right)$ that are asymptotically locally  AdS$_7 \times \mathds{S}^4$. These solutions are holographically dual to the 6d maximally superconformal field theory with 2d superconformal defects invariant under $\mathfrak{d}\left(2,1;\gamma\right)\oplus \mathfrak{d}\left(2,1;\gamma\right)$. We show that a limit of these solutions, in which $\gamma \to -\infty$, reproduces another known class of solutions, holographically dual to \emph{small} $\N=(4,4)$ superconformal defects. We then use this limit to generate new small $\N=(4,4)$ solutions with finite Ricci scalar, in contrast to the known small $\N=(4,4)$ solutions. We then use holography to compute the entanglement entropy of a spherical region centered on these small $\N=(4,4)$ defects, which provides a linear combination of defect Weyl anomaly coefficients that characterizes the number of defect-localized degrees of freedom. We also comment on the generalization of our results to include $\mathcal{N}=(0,4)$ surface defects through orbifolding.}

\arxivnumber{2402.11745}

\begin{document}

\maketitle

\section{Introduction}
\label{sec:intro}

Six dimensional (6d) superconformal field theories (SCFTs) hold a special place among quantum field theories (QFTs). Owing to the classification discovered in the seminal work by Nahm \cite{Nahm:1977tg}, superconformal symmetry is only possible in six and fewer spacetime dimensions. Moreover, $\N=(2,0)$ is the maximal amount of supersymmetry (SUSY) that a 6d theory can have.  Combining this amount of SUSY with conformal symmetry constrains a 6d $\N=(2,0)$ SCFT to such a degree that the only additional information that is necessary to completely determine the theory is the choice of a gauge algebra. 

The study of the 6d $\N=(2,0)$ theory is thus of fundamental importance in QFT, for many reasons. For example, the 6d $\N=(2,0)$ SCFT determines the physics of many other QFTs in 6d, via SUSY-breaking deformations \cite{Antoniadis:1998ki,Antoniadis:1998ep,Cordova:2016xhm} such as orbifolds \cite{Blum:1997fw,Blum:1997mm,Brunner:1997gf,Intriligator:1997dh}. 
By suitable (partial) topological twisting, the 6d theory compactified on, e.g., a Riemann surface \cite{Gaiotto:2009we} or a 3-manifold \cite{Dimofte:2011ju} can also determine the physics of infinite families of QFTs in $d<6$.

The 6d $\N=(2,0)$ SCFT is also of fundamental importance in quantum gravity. Currently, the leading candidate for an ultra-violet (UV)-complete theory of quantum gravity is M-theory. M-theory's fundamental objects are M2-branes \cite{Duff:1990xz} and M5-branes \cite{Gueven:1992hh}, and the low-energy worldvolume theory on $M$ coincident M5-branes is the 6d $\N=(2,0)$ SCFT with gauge algebra $A_{M-1}$ \cite{Strominger:1995ac}. Understanding the 6d $\N=(2,0)$ SCFT is thus essential to understanding M-theory in general. In particular, via the Anti-de Sitter/CFT (AdS/CFT) correspondence, the 6d $\N=(2,0)$ SCFT can provide a fully non-perturbative definition of M-theory on an asymptotically 7d AdS spacetime, AdS$_7$, times a four-sphere, $\mathds{S}^4$ \cite{ Witten:1996hc, Maldacena:1997re, Witten:1998wy}.

Strongly interacting SCFTs constructed in string- and M-theory, including the non-Abelian 6d $\N=(2,0)$ SCFT, are prohibitively difficult to study, for many reasons, of which we will mention only three. First, the $\N=(1,0)$ and $\N=(2,0)$ SUSY multiplets include a chiral two-form gauge field, and writing a local, gauge-invariant Lagrangian for a non-Abelian higher-form gauge field remains a major open problem. These SCFTs thus have no known Lagrangian descriptions. Second, in the space of renormalization group (RG) flows, these SCFTs are \textit{isolated} fixed points, and in particular they cannot be reached as infra-red (IR) fixed points of RG flows from free ultra-violet (UV) fixed points. Third, these SCFTs are intrinsically strongly interacting. For example, the 6d $\N=(2,0)$ SCFT has no dimensionless parameter besides $M$ that can be tuned to allow a perturbative expansion.

As a result, practically all of our direct knowledge\footnote{Indirect methods such as dimensional reduction to 5d $\N\leq 2$ SUSY QFTs have also been used to great effect to study these theories, e.g. by using the resulting lower dimensional Lagrangian description together with supersymmetric localization techniques \cite{Bullimore:2014upa}.} of interacting 6d SCFTs comes from non-perturbative methods, such as the superconformal bootstrap \cite{Beem:2015aoa}, F-theory \cite{Heckman:2015bfa}, and especially AdS/CFT \cite{Apruzzi:2013yva}, where holographic computations of quantities like Weyl anomalies \cite{Henningson:1998gx} and entanglement entropy (EE) \cite{Hung:2011xb} are used to great effect to characterize 6d SCFTs at large $M$.

 An aspect of 6d SCFTs, and generally of QFTs in three and higher dimensions, that requires particularly careful treatment to characterize is the spectrum of 2d, string-like or surface, defects. In the co-dimension one case, 2d defects in 3d QFTs arise as boundaries or interfaces, and so are more easily studied and, thus, more familiar than their higher co-dimension realizations.  Despite being somewhat more exotic in standard treatments of QFTs, 2d defects of co-dimension two and greater show up in a number of settings\footnote{This is by no means an exhaustive list of the work done on 2d defects. For a recent review of boundaries and defects in QFTs and further references on the topic, see \cite{Andrei:2018die}.}: from free field theories \cite{Soderberg:2017oaa, Lauria:2020emq,Giombi:2021uae, Bianchi:2021snj} to strongly interacting and non-Lagrangian 4d QFTs, e.g. \cite{Gukov:2006jk, Gukov:2008sn, Alday:2009fs}, to being fundamental objects in 6d SCFTs \cite{Ganor:1996nf} and in the study of EE and Renyi entropies \cite{Hung:2014npa, Lewkowycz:2014jia}.  As such, the last few decades have seen tremendous advancements in characterizing \cite{Billo:2016cpy} and constraining \cite{Jensen:2015swa, Kobayashi:2018lil} the properties of 2d defects, and it is vitally important in the study of QFTs, generally, to continue this effort by finding novel constructions of surface defects and examining their unique physics.

Of interest to us in the current work are the holographic descriptions, afforded by AdS/CFT, of 6d SCFTs and the defects that they support.  In particular, we will primarily focus our attention on solutions to 11d supergravity (SUGRA) that are contained in a one-parameter family of solutions with superisometry given by the exceptional Lie superalgebra $\mathfrak{d}(2,1;\gamma)\oplus\mathfrak{d}(2,1;\gamma)$ \cite{Bachas:2013vza}. Crucially, an asymptotically AdS$_7\times \mathds{S}^4$ solution is possible only at certain values of $\gamma$. Indeed, as a historical note, prior to the full classification given in \cite{Bachas:2013vza}, evidence of $\mathfrak{d}(2,1;\gamma)\oplus\mathfrak{d}(2,1;\gamma)$ invariant SUGRA solutions that exist for general $\gamma$ were found in superconformal Janus solutions in $4d$ gauged $\mathcal{N}=8$ SUGRA \cite{Bobev:2013yra}, which extended beyond the known $\gamma=1$ AdS$_4\times\mathds{S}^7$ Janus solution of 11d SUGRA\cite{DHoker:2009lky}. 

More pertitent to the cases that we will study below, the most well studied case among the values of $\gamma$ that admit asymptotically locally AdS$_7\times \mathds{S}^4$ solutions is $\gamma=-1/2$, which holographically describes 1/2-BPS Wilson surface operators in the 6d $A_{M-1}$ $\mathcal{N}=(2,0)$ SCFT that preserve a large $\mathcal{N}=(4,4)$ 2d SUSY.  These BPS Wilson surface operators have a long history in M-theory descriptions of 6d SCFTs \cite{Strominger:1995ac,Ganor:1996nf,Howe:1997ue,Maldacena:1998im, Berenstein:1998ij}, and there has been a recent resurgence of interest in these defects where holographic \cite{Gentle:2015jma, Estes:2018tnu, Jensen:2018rxu} and field theoretic \cite{Chalabi:2020iie,Drukker:2020atp,Drukker:2020dcz} computations have characterized these defect CFTs through their EE and Weyl anomalies.

Recently, new solutions in 11d SUGRA have been constructed that are proposed to be holographically dual to 2d BPS surface defects in 6d $\mathcal{N}=(1,0)$ SCFTs preserving ``small" $\mathcal{N}=(4,4)$ or $\mathcal{N}=(0,4)$ 2d SUSY \cite{Faedo:2020nol}.  In the sections below, we will clearly demonstrate that these new solutions also fit into the one-parameter family of solutions in \cite{Bachas:2013vza} in the limit where $\gamma\to-\infty$. It will turn out that the solutions in \cite{Faedo:2020nol} are, in fact, a particular case within a broader class of solutions that can be realized in the $\gamma\to-\infty$ limit, which we will characterize by computing their contributions to the EE of a spherical region.

A crucial point that we emphasize in our construction of the $\gamma\to-\infty$ solutions is that the superisometry algebra of the near-horizon limit of a stack of M5-branes, $\mathfrak{osp}(8^*|4)$, does not contain $\mathfrak{d}\left(2,1;\gamma\right)\oplus \mathfrak{d}\left(2,1;\gamma\right)$ as a sub-superalgebra \cite{frappat1996dictionary,D_Hoker_2008,D_Hoker_2008b}.  In the supergravity solution, this is manifested by the appearance of additional terms in the four-form flux as compared to that due to pure M5-branes.  Nevertheless, the geometry is still locally asymptotically AdS$_7\times \mathds{S}^4$, which is in line with the fact that the bosonic subalgebra of $\mathfrak{d}\left(2,1;\gamma\right)\oplus \mathfrak{d}\left(2,1;\gamma\right)$ is a subalgebra of $\mathfrak{osp}(8^*|4)$.

One upshot of our analysis that follows from the identification of the global symmetries in the $\gamma\to-\infty$ limit is that it allows for a suitable regulation scheme, which we will employ when we compute the defect sphere EE.  Lacking a generalized program of holographic renormalization for SUGRA solutions dual to defects on the field theory side, we will use a background subtraction scheme in order to remove the ambient degrees of freedom and isolate contributions to physical quantities coming from the defect.  The key step in this background subtraction scheme is the correct identification of the vacuum solution, and as will be made clear, the ambient theory with a ``trivial defect'' is a deformed 6d $\N=(2,0)$ SCFT preserving the bosonic superconformal subalgebra {$\mf{so}(2,2)\oplus\mf{so}(4)\oplus\mf{so}(5)\subset \mf{osp}(8*|4)$}. Since the solutions in \cite{Faedo:2020nol} belong to the class of solutions that we study below, in finding a vacuum solution that manifests the corresponding isometries and carefully carrying out background subtraction, we will also resolve a puzzle in \cite{Faedo:2020nol}, where physical quantities like the ``defect central charge'' were divergent.

Ultimately, we will see that that universal part of the defect sphere EE, $S_{\rm EE}^{\tiny\rm (univ)}$, i.e. the coefficient of its log-divergent part, is determined in terms of the highest weight vector, $\varpi$, of an $A_{M-1}$ irreducible representation determined by a Young diagram that encodes the partition of M5-branes that specifies the defect.  Explicitly, we will show that 
\begin{align}\label{eq:EE-intro}
    S_{\rm EE}^{\tiny\rm (univ)} = - \frac{(\varpi,\varpi)}{5}~,
\end{align}
where $(\cdot,\cdot)$ is the scalar product on the weight space. This result is similar to the Wilson surface sphere EE \cite{Estes:2018tnu} in that both are expressible in terms of scalar quantities derived from representation data but differ in that \eq{EE-intro} is negative definite\footnote{Unlike in an ordinary 2d CFT, \eq{EE-intro} being strictly negative does not necessarily signal that the theory may be non-unitary.  Indeed, the 2d defect sphere EE is expressible as a signed linear combination of defect Weyl anomaly coefficients \cite{Jensen:2018rxu}, which is not bounded from below.} and is completely determined by the highest weight vector.

In \sn{review}, we begin by reviewing the 11d supergravity solutions dual to 2d small $\mathcal{N}=(4,4)$  defects in 6d $\mathcal{N}=(1,0)$ SCFTs found in \cite{Faedo:2020nol}.  In \sn{large2small} we briefly review the $\mathfrak{d}(2,1;\gamma)\oplus\mathfrak{d}(2,1;\gamma)$-invariant solutions to 11d supergravity found in \cite{Bachas:2013vza}, and we show that by orbifolding the solutions in the $\gamma\to-\infty$ limit, we can recover the solutions in \cite{Faedo:2020nol}.  We then use the $\gamma\to-\infty$ limit to construct new 2d small $\mathcal{N}=(4,4)$ defects with finite Ricci scalar in \sn{newsmall}. Further in \sn{newsmall}, we demonstrate that the na\"ive AdS$_7\times\mathds{S}^4$ vacuum is inappropriate to use in a background subtraction scheme for regulating holographic computations in the $\gamma\to-\infty$ limit, and we identify the correct background to use in this scheme.  In \sn{EE}, we utilize the new $\gamma\to-\infty$ supergravity solutions and correct regulating scheme in a computation of the contribution of a flat 2d small $\mathcal{N}=(4,4)$ superconformal defect to the EE of a spherical region in a 6d SCFT.  We then summarize our findings and discuss remaining issues and open questions surrounding these new defect solutions in \sn{Discussion}.  

In addition, there are two appendices that detail technical aspects of the computations in the main text.  First, in \app{FG}, we analyze the asymptotic expansion  of the supergravity data that specify the new solutions in the $\gamma\to-\infty$ limit and construct the map to Fefferman-Graham (FG) gauge.  Lastly, in \app{HEE-integral}, we carefully treat the integral in the area functional of the Ryu-Takayanagi (RT) surface in the computation of the holographic EE of the defect in the dual field theory.

%%%%%%%%%%%%%%%%%%%%%%%%%%%%%%%%%%%%%%%%%%%%%%%%%%
%%%%%%%%%%%%%%%%%%%%%%%%%%%%%%%%%%%%%%%%%%%%%%%%%%
\section{\texorpdfstring{\boldmath Review: small ${\N=(4,4)}$ surface defects}{Review: small N=(4,4) surface defects}}

\label{sec:review}
%%%%%%%%%%%%%%%%%%%%%%%%%%%%%%%%%%%%%%%%%%%%%%%%%%
%%%%%%%%%%%%%%%%%%%%%%%%%%%%%%%%%%%%%%%%%%%%%%%%%%

\begin{table}[t]
    \centering
    \begin{tabular}{|c||c | c c c | c | c | c c c | c|}
         \hline
         & $\mathbb{R}^{1,1}$ & $r$ & $\theta^1$ & $\theta^2$ & $\chi$ & $z$ & $\rho$ & $\varphi^1$ & $\varphi^2$ & $\phi$ \\
         \hline\hline
        KK$^\prime$ & \begin{tabular}{c c} -- & -- \end{tabular} & -- & -- & -- & -- & -- & $\cdot$ & $\cdot$ & $\cdot$ & ISO\\
        M5$^\prime$ & \begin{tabular}{c c} -- & -- \end{tabular} & -- & -- & -- & -- & $\cdot$ & $\cdot$ & $\cdot$ & $\cdot$ & $\cdot$\\
        M2 & \begin{tabular}{c c} -- & -- \end{tabular} & $\cdot$ & $\cdot$ & $\cdot$ & $\cdot$ & -- & {$\sim$} & {$\sim$} & {$\sim$} & {$\sim$}\\
        M5 & \begin{tabular}{c c} -- & -- \end{tabular} & $\cdot$ & $\cdot$ & $\cdot$ & $\cdot$ & {$\sim$} & -- & -- & -- & --\\
        KK & \begin{tabular}{c c} -- & -- \end{tabular} & $\cdot$ & $\cdot$ & $\cdot$ & ISO & -- & -- & -- & -- & --\\ \hline
    \end{tabular}
    \caption{The 1/8-BPS brane setup of \cite{Faedo:2020nol}, with M2-M5 defect branes intersecting orthogonal M5$^\prime$-branes, and with both stacks of 5-branes probing A-type singularities. In our conventions, --, $\cdot$ and $\sim$ denote directions along which a brane is extended, localized, and smeared, respectively, while ISO(metric) denotes the compact direction of the KK-monopoles.}
    \label{tab:braneintersection}
\end{table}

We begin with a brief summary of the results of \cite{Faedo:2020nol}. The particular 11d supergravity metric constructed therein is the uplift of a 7d charged AdS$_3\times\mathds{S}^3$ domain wall initially found in \cite{Dibitetto:2017tve}, and is given by
\begin{align}\label{eq:Yolanda-Near-Horizon}
    \text{d}s^2=4 k Q_{\text{M5}}H_{\text{M5}^\prime}^{-1/3}\left(\text{d}s^2_{\text{AdS}_3}+\text{d}s^2_{\mathds{S}^3/\mathbb{Z}_k}\right)+H_{\text{M5}^\prime}^{2/3}\left(\text{d}z^2+\text{d}\rho^2+\rho^2\text{d}s^2_{\tilde{\mathds{S}}^3/\mathbb{Z}_{k^\prime}}\right),
\end{align}
for some parameter $Q_{\text{M5}}$ and a function $H_{\text{M5}^\prime}$ defined over a 4d space parametrized by the coordinates $\{z,\rho,\varphi^1,\varphi^2\}$. The solution above captures the near-horizon geometry of the brane intersection depicted in \tbl{braneintersection}. Namely, a ``bound state'' (in the sense discussed in footnote \ref{foot:boundstate}) of M2- and M5-branes, with charges $Q_{\text{M2}}$ and $Q_{\text{M5}}$, intersects an orthogonal stack of M5$^\prime$-branes, thus forming a 1/4-BPS brane setup. In 2d notation, this corresponds to $\mathcal{N}=(4,4)$ supersymmetry, with the large R-symmetry realized geometrically as the isometry of the two 3-spheres $\mathds{S}^3$ and $\tilde{\mathds{S}}^3$ with coordinates $\{\chi,\theta^1,\theta^2\}$ and $\{\phi,\varphi^1,\varphi^2\}$, respectively. In addition, the two stacks of 5-branes can be made to probe ALE singularities by introducing two Kaluza-Klein (KK) monopoles, with charges $k$ and $k^\prime$ and Taub-NUT directions $\partial_\chi$ and $\partial_\phi$, respectively. The inclusion of any one of the two KK monopoles results in a further breaking of the preserved supersymmetries and a degeneration to an 1/8-BPS setup, which in 2d language corresponds to (large) $\mathcal{N}=(0,4)$ supersymmetry. The presence of the second KK monopole does not incur a further loss of supersymmetry, so the final brane configuration is always at least a $1/8$-BPS supergravity solution.

Furthermore, the defect M2- and M5-branes are fully localized within a 2d submanifold of the worldvolume of the M5$^\prime$-branes. This is to be expected from the holographic realization of a surface defect in a 6d SCFT; this interpretation was first attached to the 7d domain wall in \cite{Dibitetto:2017klx} and to the full 11d SUGRA background in \cite{Faedo:2020nol}. On the other hand, the defect branes are smeared in the directions transverse to the worldvolume of the M5$^\prime$-branes\footnote{\label{foot:boundstate}It is this property -- that the M2- and M5-branes do not share transverse directions with the M5$^\prime$-branes, other than those along which the former are smeared -- which we are implicitly referring to when we describe the M2- and M5-branes as forming a ``bound state''. This aligns with the terminology used in \cite{Faedo:2020nol}, and should not to be confused with the dyonic supermembrane \cite{Izquierdo:1995ms}, which is a rather different multimembrane solution of 11d supergravity. Indeed, in the setup described by \tbl{braneintersection}, there is no M2-brane charge dissolved within the M5-brane worldvolume, nor do the M2-branes polarize into a fuzzy 3-sphere via the Myers effect.}, so that their charge is localized within $\mathds{S}^3/\mathbb{Z}_k$, but not $\tilde{\mathds{S}}^3/\mathbb{Z}_{k^\prime}$. Therefore, while the metric in \eq{Yolanda-Near-Horizon} manifests the isometry groups of both (orbifolded) 3-spheres, the R-symmetry is partially broken, giving rise to small $\mathcal{N}=(0,4)$ supersymmetry. For $k^\prime=1$, the solution above fits into the classification of $\mathcal{N}=(0,4)$ AdS$_3\times\mathds{S}^3/\mathbb{Z}_k\times\operatorname{CY}_2$ backgrounds foliated over an interval performed in \cite{Lozano:2020bxo}, where $\operatorname{CY}_2=\mathbb{R}^4$ contains the (round) $\tilde{\mathds{S}}^3$. In particular, the solution above corresponds to taking the M5$^\prime$-branes to be completely localized in their transverse space.

On shell, the defect brane charges are constrained to be equal, $Q_{\text{M2}}=Q_{\text{M5}}$, while the function $H_{\text{M5}^\prime}$ satisfies \cite{Faedo:2020nol}
\begin{align}\label{eq:Yolanda-Equation}
    \nabla^2_{\mathbb{R}^3_{\hat{\rho}}}H_{\text{M5}^\prime}(z,\hat{\rho})+\frac{k^\prime\partial_z^2H_{\text{M5}^\prime}(z,\hat{\rho})}{\hat{\rho}}=0,
\end{align}
where we rescaled $\hat{\rho}=\rho^2/(4k^\prime)$, and denoted by $\mathbb{R}^3_{\hat{\rho}}$ the three-dimensional subspace which is transverse to the M5-branes, parallel to the M5$^\prime$-branes, and along which the M2-branes are smeared. Following the parametrization adopted in \tbl{braneintersection}, $\mathbb{R}^3_{\hat{\rho}}$ is then the space spanned by $\{\partial_{\hat{\rho}},\partial_{\varphi^1},\partial_{\varphi^2}\}$, with $\hat\rho$ being the radial coordinate and $\{\varphi^1,\varphi^2\}$ parametrizing a 2-sphere.

In terms of the brane setup described above, then, the spacetime in \eq{Yolanda-Near-Horizon} is reached by approaching the brane intersection locus from within the worldvolume of the M5$^\prime$-branes in a radial fashion, i.e. by taking $r\to0$. In this limit, the $ISO(1,1)$ isometry group gets promoted to $SO(2,2)$, and the M5$^\prime$-brane worldvolume becomes AdS$_3\times\mathds{S}^3/\mathbb{Z}_k$.

A particular solution to \eq{Yolanda-Equation} is, for any $\alpha\in\mathbb{R}$,
\begin{equation}\label{eq:Yolanda-Solution}
    H_{\text{M5}^\prime}(z,\rho) = \frac{4 \sqrt{2}}{g^3} \frac{1}{P_+ P_-} \frac{\sqrt{P_+^2 + P_-^2 -4 \alpha^2 + 2 P_+ P_-}}{P_+^2 + P_-^2 + 2 P_+ P_-},
\end{equation}
where $P_\pm=\sqrt{z^2+(\rho\pm\alpha)^2}$. By redefining
\begin{subequations}
    \begin{align}
        \rho&= \alpha \frac{\cos\xi}{\sqrt{1 - \mu^5}}\\
        z&= \alpha \mu^\frac{5}{2} \frac{\sin\xi}{\sqrt{1 - \mu^5}}
\end{align}
\end{subequations}
and setting $\alpha=(2^{7/4}g^{3/2}kQ_{\text{M5}})^{-1}$, the particular solution in \eq{Yolanda-Solution} can be matched to the one found in eq.~(2.17) of \cite{Faedo:2020nol}, which we now reproduce for clarity\footnote{To see this, it is convenient to first rationalize the product $P_+ P_-$ as $P_+ P_- = \alpha^2 \big(1 + \mu^5 - (1 - \mu^5) \cos(2\xi)\big) / 2(1-\mu^5)$.}:
\begin{equation}\label{eq:Yolanda-H}
    H_{\text{M5}^\prime}(\mu,\xi)=2^{27/4}(\sqrt{g}kQ_{\text{M5}})^{3}\frac{\mu^{5/2}(1-\mu^5)^{3/2}}{\mu^5\cos^2\xi+\sin^2\xi}~.
\end{equation}
Furthermore, in \cite{Faedo:2020nol} it was argued that, as $\mu\to 1$ (which corresponds to a non-linear limit in the original coordinates $\hat\rho$ and $z$), the near-horizon geometry in \eq{Yolanda-Near-Horizon} locally recovers the AdS$_7/\mathbb{Z}_k\times\mathds{S}^4/\mathbb{Z}_{k^\prime}$ vacuum of M-theory. This was shown by realizing the 11d line element as the uplift of the domain wall solution to $\mathcal{N}=1$, $d=7$ supergravity (whose gauge coupling constant $g$ appears in \eq{Yolanda-H} above) found in \cite{Dibitetto:2017tve}, which interpolates between AdS$_7$ (as $\mu\to1$) and an infrared singularity (at $\mu=0$).

While the singular nature of the solution is not immediately obvious, it can be made manifest by studying the Ricci scalar, $\mathcal{R}$, in the $z\to 0$ limit.  For metrics generally of the form of \eq{Yolanda-Near-Horizon}, and following \cite{D_Hoker_2008b}, we can write the expression for the Ricci scalar for an arbitrary harmonic function $H_{\text{M5}^\prime}$ as
\begin{align}\label{eq:Yolanda-Ricci}
    \mathcal{R}  &= 
    H_{\rm M5^\prime}^{-2/3} \left[ \frac{1}{6} \frac{(\partial_z H_{\rm M5^\prime})^2 + (\partial_\rho H_{\rm M5^\prime})^2 }{H_{\rm M5^\prime}^2}
    - \frac{2}{3} \frac{\partial_z^2 H_{\rm M5^\prime} + \partial_\rho^2 H_{\rm M5^\prime}}{H_{\rm M5^\prime}}
    - 2 \frac{\partial_\rho H_{\rm M5^\prime}}{\rho H_{\rm M5^\prime}} \right].
\end{align}

The function $H_{\rm M5^\prime}$ has a branch point located at $z = 0$ and $\rho = \alpha$ and correspondingly admits two different expansions as $z \rightarrow 0$, depending on whether $\rho$ is larger or smaller than $\alpha$.  The choice of sign for the branch cut can be determined by the requirement $P_+ P_- \geq 0$.  For $\rho > \alpha$, this leads to
\begin{align}
\mathcal{R} = \frac{g^2 (2 \alpha^2 - 3 \rho^2)^2}{12 \rho^{2/3} (\rho^2 - \alpha^2)^{5/3}} + \mathcal{O}(z^2)
\end{align}
which has a pole as $\rho \rightarrow \alpha$.  This corresponds to setting $\xi = 0$, with the pole appearing as $\mu \rightarrow 0$.  For $\rho < \alpha$, we find
\begin{align}
\mathcal{R} = \frac{g^2 \alpha^{2/3}(\alpha^2 - \rho^2)}{12 z^{8/3}} + \mathcal{O}\bigg(\frac{1}{z^{2/3}}\bigg)
\end{align}
which corresponds to taking $\mu \rightarrow 0$ with $\xi \neq 0$.  Thus for all values of $\xi$, we find that the Ricci scalar diverges as $\mu \rightarrow 0$.

%%%%%%%%%%%%%%%%%%%%%%%%%%%%%%%%%%%%%%%%%%%%%%%%%%
%%%%%%%%%%%%%%%%%%%%%%%%%%%%%%%%%%%%%%%%%%%%%%%%%%
\section{\texorpdfstring{\boldmath From large to small $\N=(4,4)$ surface defects}{From large to small N=(4,4) surface defects}}
\label{sec:large2small}
%%%%%%%%%%%%%%%%%%%%%%%%%%%%%%%%%%%%%%%%%%%%%%%%%%
%%%%%%%%%%%%%%%%%%%%%%%%%%%%%%%%%%%%%%%%%%%%%%%%%%

To begin, we will briefly review the classification due to \cite{D_Hoker_2008b,Bachas:2013vza} of the $d=11$ supergravity solutions with superisometry given by two copies of the exceptional Lie superalgebra $\mf{d}(2,1;\gamma)$. This classification represents the foundation for the new defect solutions we present below. 

In particular, we will be interested in the real form $\mathfrak{d}(2,1;\gamma;0)$ which arises as a real subsuperalgebra of the fixed points of an involutive semimorphism of $\mathfrak{d}(2,1;\gamma)$ \cite{frappat1996dictionary}. Recall that the 9-dimensional maximal bosonic subalgebra of the real form $\mf{d}(2,1;\gamma;0)$ is $\mf{so}(2,1)\oplus\mf{so}(3)\oplus\mf{so}(3)$.  Labelling each factor in this subalgebra by an index $a\in\{1,2,3\}$, the generators $T^{(a)}$ satisfy
\begin{align}
    [T^{(a)}_i,\, T^{(b)}_j]  = i\delta^{ab}\varepsilon_{ijk}\eta^{kl}_{a}T^{(a)}_l\quad\text{for }i,j\in\{1,2,3\}~,
\end{align}
where $\varepsilon_{ijk}$ and $\eta^{kl}_{a}$  are the totally anti-symmetric tensor (with $\varepsilon_{123}=1$) and the canonical metric induced by the Killing form, respectively. In addition to the bosonic sector, $\mf{d}(2,1;\gamma)$ contains an 8-dimensional fermionic generator with components $F_{A_1 A_2 A_3}$, where the indices $A_a\in\{\pm\}$ transform in the spinorial representation $\boldsymbol{2}$ of the $a^{\rm th}$ factor in the even subalgebra.  Furthermore, the components $F_{A_1 A_2 A_3}$ obey the following anti-commutation relation
\begin{align}\label{eq:anticommutator}
\{ F_{A_1 A_2 A_3},F_{B_1 B_2 B_3} \} =\ & \beta_1 C_{A_2 B_2} C_{A_3 B_3} (C \sigma^i)_{A_1 B_1} T_i^{(1)} \\\nonumber
& +\beta_2 C_{A_1 B_1} C_{A_3 B_3} (C \sigma^i)_{A_2 B_2} T_i^{(2)} \\\nonumber
& +\beta_3 C_{A_1 B_1} C_{A_2 B_2} (C \sigma^i)_{A_3 B_3} T_i^{(3)}~,
\end{align}
where $C\equiv i \sigma^2$ is the charge conjugation matrix, $\{\sigma^i\}_{i=1}^3$ are the Pauli matrices, and $\{\beta_i\}_{i=1}^3$ are real parameters satisfying $\sum_{a=1}^3 \beta_a =0$, which follows from the (generalized) Jacobi identity.  This last constraint, together with the possibility of absorbing any rescaling $\{\lambda\beta_1,\lambda\beta_2,\lambda\beta_3\}$ for $\lambda\in\mathbb{C}$ into a redefinition of the normalization of the fermionic generator, implies that $\mf{d}(2,1;\gamma)$ is entirely specified by the choice of a ratio of any two of the three $\beta$ parameters; here, we take $\gamma \equiv \beta_2/\beta_3$. Note that $\mf{d}(2,1;\gamma)$ is the only (finite-dimensional) Lie superalgebra admitting a continuous parametrization.

Amongst the possible values that $\gamma$ can take, there are clearly three special values corresponding to the vanishing of any one of the $\beta$ parameters.  Specifically, choosing $\beta_1=0$ fixes $\gamma=-1$. The more interesting case, and the one relevant to our analysis, is $\beta_3=0$, which corresponds to $\gamma \to \pm\infty$.  The case $\beta_2=0$ corresponds to $\gamma=0$ and is equivalent under a group involution as discussed at the end of this section. In the limit $\beta_3=0$, the anticommutator in \eq{anticommutator} degenerates to
\begin{equation}
    \{ F_{A_1 A_2 A_3},F_{B_1 B_2 B_3} \} = \beta_1\left(C_{A_2 B_2} C_{A_3 B_3} (C \sigma^i)_{A_1 B_1} T_i^{(1)}-C_{A_1 B_1} C_{A_3 B_3} (C \sigma^i)_{A_2 B_2} T_i^{(2)}\right)~.
\end{equation}
Consequently, the $\mathfrak{so}(4)_R\cong\mathfrak{so}(3)\oplus\mathfrak{so}(3)$ R-symmetry of the large superalgebra is broken into the single $\mathfrak{so}(3)_R$ factor which constitutes the R-symmetry of a small superalgebra. Note that the other $\mathfrak{so}(3)$ factor remains a bosonic symmetry of the supergravity solution; however, it is now realized as a flavor symmetry, rather than as an outer automorphism of the supersymmetry algebra.

In addition, there are isolated points of interest in the $\gamma$-parameter space where the real form $\mathfrak{d}(2,1;\gamma;0)$ reduces to classical Lie superalgebras:
\begin{subequations}\label{eq:gamma-superalgebras}
    \begin{align}
        \mathfrak{d}(2,1;\gamma;0)&=\mathfrak{osp}(4^*|2)\quad &&\text{for }\gamma\in\{-2,-1/2\}\\
        \mathfrak{d}(2,1;\gamma;0)&=\mathfrak{osp}(4|2;\mathbb{R})\quad &&\text{for }\gamma=1~.
    \end{align}
    \end{subequations}
In particular, $\gamma=-1/2$ is the only value (up to algebra involutility, as we will discuss shortly) for which $\mathfrak{d}(2,1;\gamma;0)\oplus\mathfrak{d}(2,1;\gamma;0)$ admits a canonical inclusion into the superisometry algebra $\mathfrak{osp}(8^*|4)$ of AdS$_7\times\mathds{S}^4$. This case was studied extensively in \cite{Estes:2018tnu}; it is the holographic realization of Wilson surfaces, 1/2-BPS codimension-4 superconformal solitons, within the 6d $\mathcal{N}=(2,0)$ SCFT. In this case, the ambient 6d SCFT is undeformed, in the sense that the supergroup of symmetries preserved by the defect is a subgroup of the 6d superconformal symmetry group. This is a common feature throughout the study of defects embedded in higher-dimensional theories.

For generic values of $\gamma$, including the limit $\gamma\to\pm\infty$, the superisometry algebra $\mathfrak{d}(2,1;\gamma;0)\oplus\mathfrak{d}(2,1;\gamma;0)$ is not a subalgebra of the $\mathfrak{osp}(8^*|4)$ superconformal Lie superalgebra \cite{Estes:2018tnu}. As a result, we expect the 6d ambient theory to be deformed as we tune $\gamma$ away from the special value $\gamma=-1/2$.  In the supergravity solutions we construct below, this deformation will appear at leading order in the asymptotic expansion of the four-form flux. Additionally, the warp factors of the symmetric spaces corresponding to $T^{(1)}$ and $T^{(2)}$ have the correct normalization to create an AdS$_7$ space only when $|\beta_1| = |\beta_2|$, which can happen only when $\gamma = -1/2$ or $\gamma\to\pm\infty$.

Finally, we note that the complex Lie superalgebra $\mathfrak{d}(2,1;\gamma)$ enjoys a triality symmetry generated by $\gamma\mapsto\gamma^{-1}$, $\gamma\mapsto-(\gamma+1)$, and $\gamma\mapsto-\gamma/(\gamma+1)$, any two of which are linearly independent. However, only the $\gamma\mapsto\gamma^{-1}$ involutility survives at the level of the real form $\mathfrak{d}(2,1;\gamma;0)$, due to the distinguished nature of the $T^{(1)}$ generator. Therefore, our analysis below can be equivalently carried out in the $\gamma\to\pm\infty$ limits, or even in the $\gamma\to0$ limit (albeit with a permutation of the two $\mf{so}(3)\oplus\mf{so}(3)$ factors contained in two copies of $\mf{d}(2,1;\gamma;0)$)\footnote{In spite of the involution relating the $\gamma\to- \infty$ and $\gamma\to0$ limits, these two descriptions cannot be smoothly deformed into one another by varying $\gamma$. Indeed, the two regimes are separated by the decompactification point $\gamma =-1$.}.  The physical interpretation of the choice of either limit corresponds to fixing the orientation of the M5-branes that engineer the ambient 6d SCFT. 

\subsection{\texorpdfstring{Supergravity solutions for general $\gamma$}{Supergravity solutions for general γ}}

In this section, we will review the structure of supergravity solutions with $\mf{d}(2,1;\gamma;0)\oplus \mf{d}(2,1;\gamma;0)$ superisometry algebra, for generic $\gamma$, as first described in \cite{D_Hoker_2008b,Estes:2012vm,Bachas:2013vza}. The odd subspace of this superalgebra is 16-dimensional for all $\gamma$, so that all supergravity backgrounds discussed below are $1/2$-BPS. Furthermore, the maximal bosonic subalgebra of $\mf{d}(2,1;\gamma;0)\oplus \mf{d}(2,1;\gamma;0)$ is \begin{equation}
\mathfrak{so}(2,2)\oplus \mathfrak{so}(4)\oplus\mathfrak{so}(4)~.
\end{equation}
To realize this superisometry, the supergravity metric must be of the form\footnote{For the previously mentioned special values $\gamma=-1$ and $\gamma=0$, AdS$_3$ Wigner-\.{I}n\"{o}n\"{u} contracts to $\mathbb{R}^{2,1}$ and one of the 3-spheres decompactifies into $\mathbb{R}^3$, respectively.}
\begin{equation}
(\text{AdS}_3\times\mathds{S}^3\times\tilde{\mathds{S}}^3)\ltimes\Sigma_2~,
\end{equation}
for a Riemann surface $\Sigma_2$.

The Killing spinor equations on such manifolds can be recast into conditions on two auxiliary functions, $h$ and $G$, defined over the Riemann surface $\Sigma_2$ \cite{D_Hoker_2008b,Estes:2012vm,Bachas:2013vza}. Specifically, if we employ local complex\footnote{The SUGRA orientation and Riemannian metric automatically endow the Riemann surface $\Sigma_2$ with a complex structure.} coordinates $\{w,\bar{w}\}$ on $\Sigma_2$, so that the metric on the Riemann surface is given by $\text{d}s^2_{\Sigma_2}=\text{d}w\text{d}\bar{w}$, the function $h$ is constrained to be $\mathbb{R}$-valued and harmonic
\begin{equation}
    \partial_w\partial_{\bar w}h=0~,
\end{equation}
while $G$ is $\mathbb{C}$-valued and satisfies the conformally covariant equation
\begin{equation}\label{eq:Equation-for-G}
    \partial_w G=\frac{1}{2}\left(G+\bar G\right)\partial_w\ln h~.
\end{equation}
Note that both constraints above hold for generic $\gamma$.

Regularity conditions on the auxiliary functions $h$ and $G$ have been discussed in \cite{Bachas:2013vza}. Here, we will consider Riemann surfaces with a boundary, $\partial\Sigma_2\neq \emptyset$. In short, regularity of the supergravity solution constrains $G=\pm i$ on $\partial\Sigma_2$, while the (holomorphic part of the) function $h$ must either vanish or have a simple pole at any point of $\partial\Sigma_2$.

The triple $(h,G,\gamma)$ uniquely specifies a bosonic background within the space of 11d supergravity solutions with superisometry algebra $\mf{d}(2,1;\gamma;0)\oplus \mf{d}(2,1;\gamma;0)$. In particular, the metric field in the supergravity solution can be written in terms of this data as
\begin{equation}\label{eq:Johns-Solutions}
\text{d}s^2 = f_{\text{AdS}_3}^2 \text{d}s^2_{\text{AdS}_3} + f_{\mathds{S}^3
}^2 \text{d}s^2_{\mathds{S}^3
}+ f_{\tilde{\mathds{S}}^3
}^2 \text{d}s^2_{\tilde{\mathds{S}}^3
} + f_{\Sigma_2}^2\text{d}s^2_{\Sigma_2}~,
\end{equation}
where the metric factors
are functions over $\Sigma_2$ and can be expressed in terms of the auxiliary functions 
    \begin{align}\label{eq:Auxiliary-W-pm}
        W_\pm := |G\pm i|^2 +\gamma^{\pm1}(G\bar{G}-1)
        \end{align}
as \cite{Bachas:2013vza}
\begin{subequations}\label{eq:Metric-Functions}
\begin{align}
    f_{\text{AdS}_3}^6 &= \frac{h^2 W_+ W_-}{\beta_1^6 (G \bar G - 1)^2}~, \\
    f_{{\mathds{S}}^3}^6 &= \frac{h^2 (G \bar G -1) W_-}{\beta_2^3 \beta_3^3 W_+^2}~, \\
    f_{\tilde{\mathds{S}}^3}^6 &= \frac{h^2 (G \bar G -1) W_+}{\beta_2^3 \beta_3^3 W_-^2}~, \\
    f_{\Sigma_2}^6 &= \frac{|\partial_w h|^6}{\beta_2^3 \beta_3^3 h^4} (G \bar G - 1) W_+ W_-~.
\end{align}
\end{subequations}

The bosonic sector of the 11d supergravity solution is completed by a three-form gauge potential $\mathcal{C}_{(3)}$, which can be written in terms of the $(h,G,\gamma)$ data as
\begin{align}\label{eq:C3}
\mathcal{C}_{(3)} =\sum_{i=1}^3 b_{\mathcal{M}_i}\operatorname{vol}_{\mathcal{M}_i}
\end{align}
where $\operatorname{vol}_{\mathcal{M}_i}$ denotes the volume form on the manifold $\mathcal{M}_i=\{\text{AdS}_3,\mathds{S}^3,\tilde{\mathds{S}}^3\}_i$. We also introduced the gauge potentials
\begin{subequations}\label{eq:Potentials}
    \begin{align}
        b_{\text{AdS}_3}&:=\frac{\tau_1}{\beta_1^3}\left[-\frac{h(G+\bar{G})}{1-G \bar{G}}+(2+\gamma+\gamma^{-1})\Phi-(\gamma-\gamma^{-1})\tilde{h}+b_1^0\right]~,\\
        b_{\mathds{S}^3}&:=\frac{\tau_2}{\beta_2^3}\left[-\frac{\gamma h(G+\bar{G})}{W_+}+\gamma(\Phi-\tilde{h})+b_2^0\right]~,&\\
        b_{\tilde{\mathds{S}}^3}&:=\frac{\tau_3}{\beta_3^3}\left[\frac{h(G+\bar{G})}{\gamma W_-}-\frac{\Phi+\tilde{h}}{\gamma}+b_3^0\right]~,&
    \end{align}
\end{subequations}
where $\{b^0_i\}_{i=1}^3$ are integration constants, while the dual harmonic function $\tilde{h}$ and the real auxiliary function $\Phi$ are defined in terms of $h$ and $G$ via
\begin{subequations}
    \begin{align}
    \partial_w\tilde{h}&=-i\partial_wh~,\\
    \partial_w\Phi&=\bar{G}\partial_wh~.
    \end{align}
\end{subequations}
Finally, in \eq{Potentials} we made use of the signs $\tau_i=\pm1$ for $i\in\{1,2,3\}$, which are subject to the constraint 
\begin{equation}\label{eq:sign-constraint}
    \prod_{i=1}^3\beta_if_{\mathcal{M}_i}+h\prod_{i=1}^3\tau_i=0~.
\end{equation}

As discussed above, regularity conditions force $h$ and $G$ to have prescribed behavior on $\partial\Sigma_2$. The solutions that we are interested in studying are asymptotically AdS$_7\times\mathds{S}^4$, which are indeed described by $G=\pm i$ and $h$ having a single simple pole, which corresponds to having a single asymptotic AdS$_7\times\mathds{S}^4$ region. Generically at any point on $\partial \Sigma_2$, the regularity constraint on $h$ implies that its Laurent expansion reduces to
\begin{equation}\label{eq:Harmonic-Function}
    h=-ih_0w+\operatorname{c.c.}=\frac{2h_0\sin\vartheta}{\varrho}~,
\end{equation}
for some real constant $h_0$.  In the second equality, we have introduced new convenient set of coordinates $\{\varrho,\vartheta\}$ on $\Sigma_2$ defined by $w=e^{i\vartheta}/\varrho$. Adopting these new polar coordinates and using \eq{Harmonic-Function} allows us to perturbatively solve \eq{Equation-for-G} in small $\varrho$ to find 
\begin{equation}\label{eq:G-perturbative}
    G=-i+a_1 \varrho e^{i\vartheta}\sin\vartheta+\mathcal{O}(\varrho^2)
\end{equation}
for some constant $a_1$. If we then insert \eq{G-perturbative} into eqs.~(\ref{eq:Auxiliary-W-pm}) and (\ref{eq:Metric-Expansion}), we can perturbatively expand \eq{Johns-Solutions} in small $\varrho$ to find
\begin{equation}\label{eq:Metric-Expansion}
\text{d}s^2 = L^2 \frac{\text{d}\varrho^2}{\varrho^2}-\frac{2 \gamma L^2}{a_1 (1 + \gamma)^2\varrho}  \left( \text{d}s^2_{\text{AdS}_3} + \frac{(1 + \gamma)^2}{\gamma^2} \text{d}s^2_{\mathds{S}^3%/\mathbb{Z}_k
} \right)
+ L^2 \left(\text{d}\vartheta^2 + \sin^2\vartheta \text{d}s^2_{\tilde{\mathds{S}}^3%/\mathbb{Z}_{k'}
} \right)+\dots
\end{equation}
with
\begin{equation}
L^6 = \frac{a_1^2 h_0^2 (1+\gamma)^6}{\beta_1^6 \gamma^2}~.
\end{equation}

We can see the AdS$_7\times\mathds{S}^4$ asymptotic geometry in \eq{Metric-Expansion} clearly for certain values of $\gamma$. The obvious case is if we choose $\gamma=-1/2$, which was studied extensively in \cite{Bachas:2013vza,Gentle:2015jma,Estes:2018tnu}. The other possibility, namely the limits $\gamma\to\pm\infty$, will be discussed in the next section.

For $\gamma<0$ and for the function $h$ given in \eq{Harmonic-Function}, the general solution to \eq{Equation-for-G} for the function $G$ with an even number $2n+2$ of branch points $\{\xi_i\}_{i=1}^{2n+2}\subset\partial\Sigma_2$ was found in \cite{DHoker:2008rje} to be
\begin{equation}
\label{eq:Gwilson}
G = -i \left(1 + \sum_{j=1}^{2n+2} (-1)^j \frac{w - \xi_j}{|w - \xi_j|} \right)~,
\end{equation}
where $G$ flips sign $\pm i\to \mp i$ upon crossing each branch point $\xi_i$.

\subsection{\texorpdfstring{The $\gamma\to-\infty$ limit}{The γ → -∞ limit}}\label{sec:Gamma-To-Infinity}

Of particular interest to us in the following sections are solutions that are constructed by taking the scaling limit
\begin{align}\label{eq:Limit}
    &\gamma\to\pm\infty~,& &\gamma a_1\to\text{constant}~,& &L\to\text{constant}~.&
\end{align}
From \eq{Metric-Expansion} and the requirement $\gamma a_1\to\text{constant}$, we can see that this scaling limit realizes an AdS$_7\times\mathds{S}^4$ asymptotic geometry.  In this and the following subsections, we will consider the $\gamma\to-\infty$ limit in greater detail and show that the solutions engineered in this limit contain the class of solutions found in \cite{Faedo:2020nol} which we reviewed in \sn{review}. Below, we will expand upon these solutions and construct new surface defects with small $\mathcal{N}=(4,4)$ SUSY.

To begin, we introduce a complex function $F$ via
\begin{equation}
G = -i \left(1 + \gamma^{-1} F \right)~,
\end{equation}
and we rescale $\beta_1=\hat{\beta}(-\gamma)^{1/3}$. This ensures that $L$ remains finite as required by the limiting procedure of \eq{Limit}. In this limit, the metric in \eq{Metric-Expansion} becomes
\begin{align}\label{eq:Metric-Gamma-Limit}
\text{d}s^2 &= \left[ \frac{4 h^2}{\hat{\beta}^6 (F + \bar F)} \right]^\frac{1}{3} \left( \text{d}s^2_{\text{AdS}_3} + \text{d}s^2_{\mathds{S}^3%/\mathbb{Z}_k
} \right)
+ \left[ \frac{h^2 (F + \bar F)^2}{16 \hat{\beta}^6} \right]^\frac{1}{3} \left(\text{d}s^2_{\tilde{\mathds{S}}^3%/\mathbb{Z}_{k'}
}+ \frac{4|\partial_w h|^2}{h^2} \text{d}w\text{d}\bar w\right)~.
\end{align}
In addition, the gauge potentials are given by
\begin{align}
b_{\text{AdS}_3} = b_{\mathds{S}^3} = - \frac{2 \tilde h}{\hat{\beta}^3},\qquad
b_{\tilde{\mathds{S}}^3} = \frac{h}{2 \hat{\beta}^3} \frac{-i(F - \bar F)}{2} - \hat \Phi~,
\end{align}
where the function $\hat\Phi$ is defined via
\begin{align}
\partial_w \hat \Phi &= \bar F \frac{\partial_w \tilde h}{\hat{\beta}^3}~.
\end{align}

Let us introduce a new set of coordinates $\{z,\rho\}$ on $\Sigma_2$ defined via $w=z+i\rho$, and choose $h_0$ in \eq{Harmonic-Function} such that the harmonic function $h=\hat{\beta}^3h_1\rho$ for some constant $h_1$. Let us also parametrize the real and imaginary parts of the complex function $F$ such that
\begin{equation}
F = \frac{2 \rho^2}{h_1} H + i F_I~,
\end{equation}
for real functions $H$ and $F_I$. Note that the latter does not enter the metric field; indeed, we can rewrite the line element in \eq{Metric-Gamma-Limit} entirely in terms of $H$ as
\begin{equation}\label{eq:metric}
\text{d}s^2 = h_1H^{-\frac{1}{3}} \left( \text{d}s^2_{\text{AdS}_3} + \text{d}s^2_{\mathds{S}^3%/\mathbb{Z}_k
} \right)
+ H^\frac{2}{3} \left(\text{d}z^2 + \text{d} \rho^2 +\rho^2 \text{d}s^2_{\tilde{\mathds{S}}^3%/\mathbb{Z}_{k'}
}\right)~.
\end{equation}
The condition in \eq{Equation-for-G} implies that the complex function $F$ satisfies the first order equation
\begin{equation}
    \partial_w F = \frac{1}{2} (F - \bar F) \partial_w \ln h~,
\end{equation}
which in turn can be recast into a second order equation for its real part,
\begin{equation}\label{eq:H-Equation}
\partial_\rho^2 H + \frac{3}{\rho} \partial_\rho H + \partial_z^2 H = 0~.
\end{equation}
The supergravity solution in the $\{z,\rho\}$ parametrization of $\Sigma_2$ is completed by the following expressions for the gauge potentials,
\begin{align}
b_{\text{AdS}_3} = b_{\mathds{S}^3} = 2 h_1 z,\qquad
b_{\tilde{\mathds{S}}^3} = \frac{h_1 \rho}{2} F_I + \hat \Phi~.
\end{align}
In particular, a more explicit form can be given for the derivatives of $b_{\tilde{\mathds{S}}^3}$ as
\begin{align}
\partial_\rho b_{\tilde{\mathds{S}}^3} = \rho^3 \partial_z H,\qquad
\partial_z b_{\tilde{\mathds{S}}^3} = -\rho^3 \partial_\rho H~.
\end{align}
These solutions have a scaling symmetry under the transformation
\begin{align}\label{eq:scalesymmetry}
    &z \rightarrow \lambda z,&
    &\rho \rightarrow \lambda \rho,&
    &H \rightarrow \lambda^{-3} H,&
    &h_1 \rightarrow \lambda^{-1} h_1,&
\end{align}
for which the metric and flux are invariant.  This transformation could be used to fix the value of $h_1$ so that the solution is uniquely given by choice of function $H$.

\subsection{Inclusion of KK-monopoles and recovering known solutions}\label{sec:orbifolds}

At the level of the local geometry described by \eq{metric}, KK-monopoles can be included in a straightforward fashion by replacing $\mathds{S}^3$ and $\tilde{\mathds{S}}^3$ with the lens spaces $\mathds{S}^3/\mathbb{Z}_k$ and $\tilde{\mathds{S}}^3/\mathbb{Z}_{k^\prime}$, respectively, where $k$ and $k^\prime$ are the orbifold charges. The lens spaces can be realized as the total spaces of circle bundles over 2-spheres, where the orbifold acts on the Hopf fiber; this amounts to the substitutions
\begin{subequations}\label{eq:monopoles}
    \begin{align}
        \text{d}s^2_{\mathds{S}^3}\to&\  \text{d}s^2_{\mathds{S}^3/\mathbb{Z}_k}\ =\frac{1}{4}\left[\left(\frac{\text{d}\chi}{k}+\omega\right)^2+\text{d}s^2_{\mathds{S}^2}\right]~, \\
        \text{d}s^2_{\tilde{\mathds{S}}^3}\to&\ \text{d}s^2_{\tilde{\mathds{S}}^3/\mathbb{Z}_{k^\prime}}=\frac{1}{4}\left[\left(\frac{\text{d}\phi}{k^\prime}+\eta\right)^2+\text{d}s^2_{\tilde{\mathds{S}}^2}\right]~,
    \end{align}
\end{subequations}
where $\text{d}\omega=\operatorname{vol}_{\mathds{S}^2}$ and $\text{d}\eta=\operatorname{vol}_{\tilde{\mathds{S}}^2}$. The inclusion of either KK-monopole incurs the loss of $1/2$ of the existing supersymmetry generators, resulting in a small $\mathcal{N}=(0,4)$ supersymmetry algebra. Indeed, dimensional reduction along the Taub-NUT direction produces a D6-brane which breaks half of the supersymmetries of the massless type IIA AdS$_7$ vacuum. The second KK-monopole can be added without bringing about any further breaking of the supersymmetries, and provides a second isometric direction upon which the background can be dimensionally reduced to massless type IIA supergravity. Therefore, the final solution is an $1/8$-BPS configuration.  At the level of the superconformal symmetry algebra, this corresponds to a reduction from $\mathfrak{d}\left(2,1;\gamma\right)\oplus \mathfrak{d}\left(2,1;\gamma\right)$ to $\mathfrak{d}\left(2,1;\gamma\right)$.

The inclusion of orbifolds allows us to match the $\gamma\to-\infty$ solutions of \sn{Gamma-To-Infinity} to those found in \cite{Faedo:2020nol} and reviewed in \sn{review}. In particular, we see that the $\gamma\to-\infty$ metric in \eq{metric}, subject to the inclusion of KK-monopoles as in \eq{monopoles}, matches the M2-M5-KK-M5$^\prime$-KK$^\prime$ near-horizon in \eq{Yolanda-Near-Horizon} under the identifications $h_1=4 k Q_{\text{M5}}$ and $H=H_{\text{M5}^\prime}$. In particular, the functions $H$ and $H_{\text{M5}^\prime}$ solve the same equation, since \eq{H-Equation} maps to \eq{Yolanda-Equation} under the rescaling $\hat\rho=\rho^2/4$. We have thus managed to fully recover the solutions described in \cite{Faedo:2020nol} as a limiting case of those in \cite{Bachas:2013vza}.

%%%%%%%%%%%%%%%%%%%%%%%%%%%%%%%%%%%%%%%%%%%%%%%%%%
%%%%%%%%%%%%%%%%%%%%%%%%%%%%%%%%%%%%%%%%%%%%%%%%%%
\section{\texorpdfstring{\boldmath New small $\N=(4,4)$ surface defects}{New small N=(4,4) surface defects}}
\label{sec:newsmall}
%%%%%%%%%%%%%%%%%%%%%%%%%%%%%%%%%%%%%%%%%%%%%%%%%%
%%%%%%%%%%%%%%%%%%%%%%%%%%%%%%%%%%%%%%%%%%%%%%%%%%

In this section, we will construct a new explicit family of solutions $H(z,\rho)$ to \eq{H-Equation}. In particular, without loss of generality we will specialize to the $\gamma\to-\infty$ limit. We recall that for $\gamma<0$, a general solution with an even number of branch points $\{\xi_i\}_{i=1}^{2n+2}\subset\partial\Sigma_2$ was given by \eq{Gwilson}. We rescale the singular points $\xi_j$ as
\begin{equation}\label{eq:Rescaling}
\xi_j=\nu_j-\gamma^{-1}\hat{\xi}_j\in\partial\Sigma_2\quad\text{for}\quad j\in\{1,2,\dots,2n+2\}~,
\end{equation}
where the collapse points satisfy $\nu_j\leq\nu_{j+1}$ for all $j$, so as to preserve the total order of the set $\{\xi_j\}$ following the $\gamma\to-\infty$ limit. Furthermore, to ensure finiteness of the complex function $F$ in this limit, we must also demand that all points $\nu_j$, for $1\leq j\leq 2n+2$, correspond to the collapse of  even-dimensional clusters $\{\xi_k,\xi_{k+1},\ldots,\xi_{k+2m+1}\}$ of singular points, where $ k\leq j\leq k+2m+1$. This can be implemented by identifying
\begin{equation}
    \nu_k\equiv\nu_{k+1}\equiv\ldots\equiv\nu_{k+2m+1}
\end{equation}
for each cluster. Finally, in order to preserve the ordering of the singular points throughout the $\gamma\to-\infty$ limit, we must also arrange the corresponding collapse parameters within each cluster so that $\hat\xi_k<\hat\xi_{k+1}<\ldots<\hat\xi_{k+2m+1}$. Without loss of generality, then, it suffices to consider a pairwise collapse of neighboring singular points, which can be realized by identifying
\begin{equation}\label{eq:Pairwise-Collapse}
\nu_{2i-1}\equiv\nu_{2i}\quad\text{for}\quad i\in\{1,2,\dots,n+1\}~.
\end{equation}
However, we will later comment on certain phenomena which appear only when the singular points collapse in clusters of 4 or more.

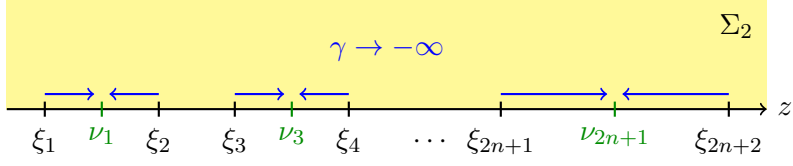
\begin{figure}[t]
    \centering
    \begin{tikzpicture}
        \fill[yellow, opacity=0.4] (-5,0) rectangle ++(10,1.5);
        \node[] at (4.6,1.1) {$\Sigma_2$};
        \draw[thick,->] (-5,0) -- (5,0) node[right]{$z$};
        \draw[thick] (-4.5,0.1) -- (-4.5,-0.1) node[below]{$\xi_1$};
        \draw[thick,black!45!green] (-3.75,0.1) -- (-3.75,-0.1) node[below]{$\nu_1$};
        \draw[thick] (-3,0.1) -- (-3,-0.1) node[below]{$\xi_2$};
        \draw[thick] (-2,0.1) -- (-2,-0.1) node[below]{$\xi_3$};
        \draw[thick,black!45!green] (-1.25,0.1) -- (-1.25,-0.1) node[below]{$\nu_3$};
        \draw[thick] (-0.5,0.1) -- (-0.5,-0.1) node[below]{$\xi_4$};

        \draw[thick] (0.6,-0.2) node[below]{$\cdots$};

        \draw[thick] (1.5,0.1) -- (1.5,-0.1) node[below]{$\xi_{2n+1}$};
        \draw[thick,black!45!green] (3,0.1) -- (3,-0.1) node[below]{$\nu_{2n+1}$};
        \draw[thick] (4.5,0.1) -- (4.5,-0.1) node[below]{$\xi_{2n+2}$};

        \draw[thick,blue,->] (-4.5,0.2) -- (-3.85,0.2);
        \draw[thick,blue,->] (-3,0.2) -- (-3.65,0.2);
        \draw[thick,blue,->] (-2,0.2) -- (-1.35,0.2);
        \draw[thick,blue,->] (-0.5,0.2) -- (-1.15,0.2);
        \draw[thick,blue,->] (1.5,0.2) -- (2.9,0.2);
        \draw[thick,blue,->] (4.5,0.2) -- (3.1,0.2);
        
        \draw[thick,blue] (-0,0.5) node[above]{$\gamma\rightarrow -\infty$};
    \end{tikzpicture}
    \caption{A particular choice of collapse points $\nu_j\in\partial\Sigma_2$ and the behavior of the singular points $\xi_i\in\partial\Sigma_2$ as $\gamma\to-\infty$ under pairwise collapse. The collapse points $\nu_j$ can be chosen arbitrarily, while maintaining $\nu_j<\nu_{j+1}$, but for clarity in the figure have been depicted at the midpoint in the region $[\xi_{j},\,\xi_{j+1}]$ to demonstrate pairwise collapse.}
    \label{fig:pairwisecollapse}
\end{figure}

An illustration of the pairwise collapse described by \eq{Rescaling} and \eq{Pairwise-Collapse} is shown in \fig{pairwisecollapse}. Under this collapse dynamic, the complex function $F$ becomes, in the $\gamma\to-\infty$ limit,
\begin{equation}
F(w,\bar{w})=\sum_{j=1}^{2n+2}(-1)^j\hat{\xi}_j\frac{\bar{w}-w}{2(\bar{w}-\nu_j)|w-\nu_j|}~.
\end{equation}
The corresponding function $H$ is given by
\begin{equation}\label{eq:H-pairwise}
H(z,\rho)=\frac{h_1}{2}\sum_{j=1}^{2n+2}\frac{(-1)^j\hat{\xi}_j}{(\rho^2+(z-\nu_j)^2)^{3/2}}~.
\end{equation}

For pure AdS$_7\times\mathds{S}^4$, this procedure produces what we refer to as the single-pole vacuum (1PV) solution with
\begin{subequations}\label{eq:1PV}
\begin{align}
F_{\text{1PV}}(w,\bar{w})&= \hat \xi \frac{\bar w - w}{\bar w |w|}~,\\
H_{\text{1PV}}(z,\rho)&= \frac{h_1 \hat \xi}{(z^2 + \rho^2)^\frac{3}{2}}~,
\end{align}
\end{subequations}
which can be obtained by taking $n=0$, $\xi_2 = -\xi_1=\xi$, and $\nu_1 = \nu_ 2 = 0$.  Alternatively, this solution can also be obtained by collapsing all singular points to the origin, i.e. $\nu_j=0$ for all $j$.
Note that \eq{H-Equation} is linear in $H$ and admits a translation symmetry under shifts of $z$.  Using these two properties, the general solutions given in \eq{H-pairwise} can be reconstructed from the 1PV solution by taking linear combinations and making use of the translation symmetry.

One may worry that since the solution for the potential $H$ in \eq{H-pairwise} is generically singular at the points $(z,\rho)=(\nu_i,0)$ for all $i\in\{1,2,\dots,2n+2\}$, the resulting supergravity metric may have a singularity or these points may simply correspond to horizons.  In order to investigate the regularity of the spacetime geometry, we compute the scalar curvature at these points.  Without loss of generality we can choose to evaluate the Ricci scalar at the point $(z,\rho)=(\nu_{2j},0)$. Recall that for metrics generally of the form in \eq{metric}, the scalar curvature is given by \eq{Yolanda-Ricci} with $H_{\rm M5^\prime}$ replaced by $ H$.
Note that since all collapse points $\nu_{2k-1}$ with odd labels are identified with even-labelled ones $\nu_{2k}$ via \eq{Pairwise-Collapse}, the point $(z,\rho)=(\nu_{2j},0)$ is indeed generic within the set of collapsed points. Ultimately, we find 
\begin{equation}\label{eq:Ricci-At-Nu}
    \mathcal{R}\bigr|_{(z,\rho)=(\nu_{2j},0)}=\frac{3}{2L_{\mathds{S}^4}^2}\left(\frac{\hat\xi_{2j}-\hat\xi_{2j-1}}{\hat{m}_1}\right)^{-2/3}~,
\end{equation}
which is indeed finite, where $L_{\mathbb{S}^4}$ and $\hat{m}_1$ are strictly positive constants introduced below.  This suggests that the geometry is regular at these points.

We also note that the 1PV solution corresponds to a spacetime with a constant scalar curvature given by
\begin{equation}
    \mathcal{R} = \frac{3}{2 L_{\mathds{S}^4}^2}~,
\end{equation}
with $L_{\mathds{S}^4} = (h_1 \hat \xi)^{1/3}$.

\subsection{Asymptotic local behavior}

We will now show explicitly that the solutions described by \eq{H-pairwise} are asymptotically locally AdS$_7\times\mathds{S}^4$. As described in greater detail in \app{FG}, the Riemann surface $\Sigma_2$ admits a parametrization by Fefferman-Graham (FG) coordinates $\{v,\phi\}$ in terms of which the line element takes the following asymptotic form for small $v$,
\begin{equation}\label{eq:FG-metric}
	\text{d}s^2=\frac{4L_{\mathds{S}^4}^2}{v^2}\left[\text{d} v^2+\alpha_1\left(\text{d} s^2_{\text{AdS}_3}+\text{d} s^2_{\mathds{S}^3}\right)\right]+L_{\mathds{S}^4}^2\left[\alpha_3\text{d}\phi^2+\alpha_4\sin^2\phi\ \text{d} s^2_{\tilde{\mathds{S}}^3}\right],
\end{equation}
where the metric factors are of the form $\alpha_i=1+\mathcal{O}(v^4)$ for $i\in\{1,3,4\}$. They are given explicitly, together with the asymptotic mapping to FG coordinates on $\Sigma_2$, in \app{FG}. The asymptotic $\mathds{S}^4$ radius $L_{\mathds{S}^4}$ can be expressed in terms of the moments
\begin{equation}\label{eq:mhat-moments}
    \hat{m}_k:=\sum_{j=1}^{2n+2}(-1)^j\hat\xi_j^k
\end{equation}
as
\begin{equation}\label{eq:radius}
	L_{\mathds{S}^4}^3=\frac{h_1\hat{m}_1}{2}~.
\end{equation}
As claimed above, we may recognize within \eq{FG-metric}, at leading order in $v$, the large-$x$ limit of the line element of AdS$_7$ (with radius $2L_{\mathds{S}^4}$) written in AdS$_3$ slicing,
\begin{equation}
\text{d}s^2_{\text{AdS}_7}=4L^2_{\mathds{S}^4}\left[\text{d}x^2+\cosh^2x\  \text{d}s^2_{\text{AdS}_3}+\sinh^2x\ \text{d}s^2_{\mathds{S}^3}\right],
\end{equation}
where the coordinate normal to the AdS$_3$ foliation is related to the FG coordinate via $x=\log(2/v)$. The large-$x$ limit trivializes the relative warping $\coth^2x$ between the AdS$_3$ and the $\mathds{S}^3$ subspaces, matching the behavior seen in \eq{FG-metric}.

In the $\gamma\to-\infty$ limit, the auxiliary functions $\Phi$ and $\tilde{h}$ admit the following FG expansions,
\begin{align}
    \Phi=-\tilde{h}&=\frac{4 \hat{m}_1 \cos \phi }{v^2}+\frac{2 \hat{n}_1}{\hat{m}_1}+\frac{\hat{m}_1 \hat{n}_2-\hat{n}_1^2}{\hat{m}_1^3}\cos \phi ~v^2+\mathcal{O}(v^4)~,
\end{align}
where the moments $\hat{n}_i$ are defined in \eq{moments-n}. Using these expansions, we find that the asymptotic geometry in \eq{FG-metric} is supported by a four-form flux\footnote{In the following, we have made a choice on the signs $\tau_{1,2,3}$ in line with the constraint $\prod_{i=1}^3\tau_i=1$ which follows from evaluating \eq{sign-constraint} on the $\gamma\to-\infty$ solutions.} 
\begin{align}\label{eq:FG-flux}
    \frac{\mathcal{F}_{(4)}}{L_{\mathds{S}^4}^3}&=-\frac{16\cos\phi}{v^3}\text{d}v\wedge(\operatorname{vol}_{\mathds{S}^3}+\operatorname{vol}_{\text{AdS}_3})-\frac{8\sin\phi}{v^2}\text{d}\phi\wedge(\operatorname{vol}_{\mathds{S}^3}+\operatorname{vol}_{\text{AdS}_3})\\\nonumber
    &\quad+3\sin^3\phi\ \text{d}\phi\wedge\operatorname{vol}_{\tilde{\mathds{S}}^3}+\ 4\cos\phi\ \frac{\hat{m}_1\hat{n}_2-\hat{n}_1^2}{\hat{m}_1^4}v\ \text{d}v\wedge(\operatorname{vol}_{\mathds{S}^3}+\operatorname{vol}_{\text{AdS}_3})+\mathcal{O}(v^2).
\end{align}
The first line in the equation above manifests the deformation of the 6d ambient SCFT by the insertion of a source, as already hinted at by the superalgebra structure discussed in \sn{large2small}. This deformation takes the form of an S-wave over the internal $\tilde{\mathds{S}}^3$. It can be compared to the undeformed theory, which corresponds to $\gamma=-1/2$. In that case, the four-form field strength at the conformal boundary $v=0$ is $\mathcal{F}_{(4)}=3L^3_{\mathds{S}^4}\operatorname{vol}_{\mathds{S}^4}$, where $\mathds{S}^4$ is the internal 4-sphere spanned by $\phi$ and $\tilde{\mathds{S}}^3$. We can recognize this term as the first term in the second line of \eq{FG-flux}.

\subsection{Single-pole vacuum}\label{sec:1PV}

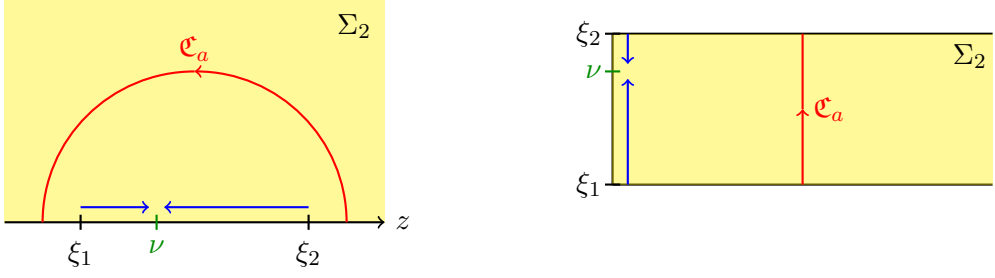
\begin{figure}[t]
    \centering
    \begin{tikzpicture}
        \fill[yellow, opacity=0.4] (-10,0) rectangle ++(5,3);
        \node[] at (-5.4,2.6) {$\Sigma_2$};
        \draw[thick,->] (-10,0) -- (-5,0) node[right]{$z$};
        \draw[thick,black!45!green] (-8,0.1) -- (-8,-0.1) node[below]{$\nu$};
        \draw[thick] (-9,0.1) -- (-9,-0.1) node[below]{$\xi_1$};
        \draw[thick] (-6,0.1) -- (-6,-0.1) node[below]{$\xi_2$};
        \draw[thick,->,blue] (-9,0.2) -- (-8.1,0.2);
        \draw[thick,->,blue] (-6,0.2) -- (-7.9,0.2);
        \draw[thick,red] (-9.5,0) arc (180:90:2) node[above]{$\mathfrak{C}_a$};
        \draw[<-,thick,red] (-7.5,2) arc (90:0:2);

        \draw[thick] (-2,0.5) -- (-2,2.5);
        \draw[thick] (-2,0.5) -- (3,0.5);
        \draw[thick] (-2,2.5) -- (3,2.5);
        \fill[yellow, opacity=0.4] (-2,0.5) rectangle ++(5,2);
        \draw[thick,red] (0.5,2.5) -- (0.5,1.5) node[right]{$\mathfrak{C}_a$};
        \draw[<-,thick,red] (0.5,1.5) -- (0.5,0.5);
        \node[] at (2.7,2.2) {$\Sigma_2$};
        
        \draw[thick] (-2.1,0.5) -- (-1.9,0.5) node[left, xshift=-1mm]{$\xi_1$};
        \draw[thick,black!45!green] (-2.1,2.0) -- (-1.9,2.0) node[left, xshift=-1mm]{$\nu$};
        \draw[thick] (-2.1,2.5) -- (-1.9,2.5) node[left, xshift=-1mm]{$\xi_2$};
        \draw[thick,->,blue] (-1.8,2.5) -- (-1.8,2.1);
        \draw[thick,->,blue] (-1.8,0.5) -- (-1.8,1.9);

    \end{tikzpicture}
    \caption{The 1PV solution described in \sn{1PV} is characterized by two singular points $\xi_{\{1,2\}}$, which collapse to the same $\nu\in\partial\Sigma_2$ in the $\gamma\to-\infty$ limit. As discussed in \sn{partition}, the basis of non-contractible four-cycles for this solution is one-dimensional. The profile of a representative cycle $\mathfrak{C}_a$ along $\Sigma_2$ is shown; note that the same 3-sphere collapses at both of its endpoints on $\partial\Sigma_2$. On the right, the 1PV on the upper half plane is mapped to a semi-infinite strip. 
    }
    \label{fig:1PVcycle}
\end{figure}

For later reference, we now identify a vacuum solution within the family of supergravity backgrounds presented above. An appropriate choice of vacuum is necessary for computing properties associated to a defect embedded in a holographic CFT. Indeed, in order to isolate quantities which are intrinsic to a defect, one must ensure that the contributions due to the ambient degrees of freedom are taken into account. The gravitational analogue of this operation, upon recasting a field theory computation into a bulk one, is vacuum subtraction. In particular we must use a vacuum solution which is characterized by the same bulk deformation as the general solutions discussed above.  As discussed above, pure AdS$_7\times\mathds{S}^4$ with $\gamma = -1/2$ does not satisfy this criteria, as the solutions we consider here with $\gamma \rightarrow -\infty$ contain a bulk deformation as can be seen in the asymptotic expression for the flux given in \eq{FG-flux}.

We take the vacuum to be the 1PV we identified in \eq{1PV}, which is the solution corresponding to the $\gamma \rightarrow -\infty$ of pure AdS$_7\times\mathds{S}^4$, i.e. two singular points $\xi_{\{1,2\}}$ collapsing to a single point $\nu$, as shown in \fig{1PVcycle}. More precisely, for general $\gamma$, the 1PV metric is given in FG form by
\begin{align}\allowdisplaybreaks
    \frac{\text{d}s^2_{\text{1PV}}(\gamma)}{L_{\mathds{S}^4}^2}=\frac{4}{v^2}&\left[\text{d}v^2+\left(1+\frac{2 \gamma +3-(2 \gamma +1) c_{2\phi}}{16 (\gamma +1)^2}v^2\right)\text{d}s^2_{\text{AdS}_3}\right.\\
    &\ +\left.\left(\frac{(\gamma +1)^2}{\gamma ^2}+\frac{2\gamma-1-(2 \gamma +1) c_{2\phi}}{16 \gamma ^2}v^2\right)\text{d}s^2_{\mathds{S}^3}+\mathcal{O}(v^4)\right]\nonumber\\
    +&\left[\left(1+\frac{(2 \gamma +1) (2 c_{2\phi}+1)}{12 (\gamma +1)^2}v^2\right)s^2_\phi\text{d}s_{\tilde{\mathds{S}}^3}^2+\left(1+\frac{(2 \gamma +1) c^2_\phi}{4 (\gamma +1)^2}v^2\right)\text{d}\phi^2+\mathcal{O}(v^4)\right], \nonumber
\end{align}
where in order to keep the expression manageable, we have adopted the abusive notation 
\begin{align}
    \cos x \equiv c_x,\quad\text{and}\quad \sin x \equiv s_x,
\end{align}
which will be employed from this point forward. For $\gamma=-1/2$, the 1PV recovers the AdS$_7\times\mathds{S}^4$ vacuum in FG gauge. In the $\gamma\to-\infty$ limit which is of relevance here, the line element of the 1PV becomes instead
\begin{equation}
    \text{d}s^2_{\text{1PV}}(\gamma\to-\infty)=\frac{4L_{\mathds{S}^4}}{v^2}\left[\text{d} v^2+\text{d} s^2_{\text{AdS}_3}+\text{d} s^2_{\mathds{S}^3} +\mathcal{O}(v^4)\right]+L_{\mathds{S}^4}^2\left[s^2_{\phi}\text{d}s_{\tilde{\mathds{S}}^3}^2+\text{d}\phi^2 +\mathcal{O}(v^4)\right].
\end{equation}
Furthermore, as a quick sanity check, we can take the 1PV limit of \eq{Ricci-At-Nu}, which recovers $\mathcal{R}|_{\rm 1PV}=3/(2L_{\mathds{S}^4}^2)$ as expected.

As remarked above, the 1PV contains no explicit defect data -- as we will see later, no Young Tableau can be associated to it. However, it does fully capture the bulk S-wave deformation discussed previously. This follows from the fact that all terms in \eq{FG-flux} which do not vanish at the conformal boundary $v=0$ are independent of the moments $\{\hat{m}_i,\hat{n}_j\}$. Therefore, the 6d ambient theory dual to the 1PV is deformed by the same sources as the theories dual to completely generic $\gamma\to-\infty$ bulk solutions, and so, in this limit, there is no smooth deformation of the field theory parameters that restores the ambient conformal symmetry in full. This is to be contrasted with the global AdS$_7\times\mathds{S}^4$ solution, which enjoys the full $SO(6,2)$ conformal symmetry. Lastly, in taking the 1PV limit, the solution exhibits a flavor symmetry enhancement $SO(4)\to SO(5)$.

In the construction of \cite{Faedo:2020nol} reviewed in \sn{review}, the 1PV solution can be obtained from \eq{Yolanda-Equation} by taking $\alpha \rightarrow 0$ and setting $g^3 = 2 \sqrt{2}/h_1 \hat \xi$.  Alternatively, it can also be obtained in the limit $g \rightarrow 0$.  To see this, first make a scale transformation using the scaling symmetry given by \eq{scalesymmetry}, take $g$ to scale with $\lambda$ as $g^3 = 2 \sqrt{2}/h_1 \hat \xi \lambda^3$, and then take $\lambda \rightarrow \infty$.

\subsection{Partition data}\label{sec:partition}

In this subsection, our aim is to identify within the $\gamma\to-\infty$ solutions of \eq{H-pairwise} a basis of independent, non-contractible cycles threaded by four-form flux. Integrating the flux along these cycles will allow us to compute the integral M-brane charges that label a supergravity solution. In turn, this characterization will enable us to recast the specification of a supergravity solution in the form of a partition containing the representation data associated to the defect string in the dual gauge theory, in analogy with the Wilson surfaces of the $\gamma = -1/2$ solutions \cite{Estes:2018tnu}. We begin by searching for independent, non-contractible four-cycles through the 11d geometry of the pairwise-collapsed $\gamma\to-\infty$ solutions. Later, we will comment on how the cycles are modified if less generic collapse dynamics are considered.

It is straightforward to see that, in the $\gamma\rightarrow-\infty$ limit, the volume of $\mathds{S}^3$ vanishes on $\partial\Sigma_2-\{\nu_{j}\}_{j=1}^{2n+2}$, i.e. all along the boundary of the Riemann surface, except at the locations of the collapse points $\nu_j$. In turn, any open curve on $\Sigma_2$ with end points on $\partial\Sigma_2-\{\nu_{j}\}_{j=1}^{2n+2}$ will be a closed curved in the full 11d geometry. Therefore, any curve encircling at least one of the distinct collapse midpoints $\nu_j$ will not be contractible to a point. This observation allows us to build a basis for non-contractible four-cycles $\mathfrak{C}_a$, by taking the product of such curves on $\Sigma_2$ with $\tilde{\mathds{S}}^3$, i.e.
\begin{equation}
    \mathfrak{C}_a\equiv\{R e^{i\theta}-\nu_{2a-1}\,|\,0\leq \theta \leq \pi\}\times \tilde{\mathds{S}}^3\,,
\end{equation}
for $1\leq a\leq n+1$. Note that any curve enveloping multiple $\nu_a$'s can be decomposed into a linear combination of curves enclosing a single collapse point; hence, to build an irreducible basis of cycles, we take the radius $R$ above such that $\mathfrak{C}_a$ encircles only a single collapse point $\nu_a$.

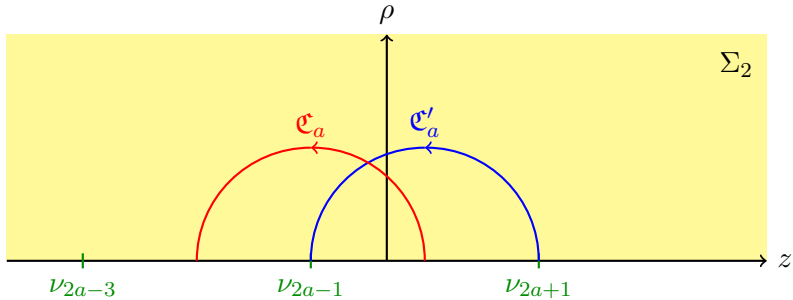
\begin{figure}[t]
    \centering
    \begin{tikzpicture}
        \fill[yellow, opacity=0.4] (-5,0) rectangle ++(10,3);
        \node[] at (4.6,2.6) {$\Sigma_2$};
        \draw[thick,->] (-5,0) -- (5,0) node[right]{$z$};
        \draw[thick,->] (0,0) -- (0,3) node[above]{$\rho$};
        \draw[thick,black!45!green] (-4,0.1) -- (-4,-0.1) node[below]{$\nu_{2a-3}$};
        \draw[thick,black!45!green] (-1,0.1) -- (-1,-0.1) node[below]{$\nu_{2a-1}$};
        \draw[thick,black!45!green] (2,0.1) -- (2,-0.1) node[below]{$\nu_{2a+1}$};
        \draw[thick,blue] (-1,0) arc (180:90:1.5) node[above]{$\mathfrak{C}^\prime_a$};
        \draw[<-,thick,blue] (0.5,1.5) arc (90:0:1.5);
        \draw[thick,red] (-2.5,0) arc (180:90:1.5) node[above]{$\mathfrak{C}_a$};
        \draw[<-,thick,red] (-1,1.5) arc (90:0:1.5);
    \end{tikzpicture}
    \caption{The profile of the non-contractible four-cycles $\mathfrak{C}_a$ and $\mathfrak{C}^\prime_a$ in $\Sigma_2$. Every point on the red and blue curves is a 3-sphere.}
    \label{fig:nonContractible4CycleBoth}
\end{figure}

The $\mathfrak{C}_a$ cycles alone do not exhaust the set of all non-contractible four-cycles. Indeed, we can consider a four-cycle constructed from a curve on $\Sigma_2$ connecting singular points $\nu_a$. For irreducibility, we only consider curves connecting neighboring collapse points. While the singular nature of their endpoints might make such curves appear problematic at first glance, building a regular four-cycle from such a curve is possible as the volume of $\mathds{S}^3$ vanishes at both endpoints, while the volume of $\tilde{\mathds{S}}^3$ remains finite. Indeed, the metric factor of $\mathds{S}^3$ contains $h_1H^{-1/3}$, while the metric factor of $\tilde{\mathds{S}}^3$ has $\rho^2H^{2/3}$, and from \eq{H-pairwise}, the function $H$ goes as $\rho^{-3}$ near any collapse point $\nu_a$. Thus, the cycle 
\begin{equation}
    \mathfrak{C}^\prime_a\equiv\left\{\frac{1}{2}(\nu_{2a+1}-\nu_{2a-1})e^{i\theta}+\frac{1}{2}(\nu_{2a+1}+\nu_{2a-1})\,\Big|\,0\leq\theta\leq\pi\right\}\times \mathds{S}^3~,
\end{equation}
has the desired behavior of a non-contractible four-cycle. The distinction between the $\mathfrak{C}_a$ and $\mathfrak{C}_a^\prime$ cycles is illustrated in \fig{nonContractible4CycleBoth}. Given that, by construction, they connect neighboring collapse points, there are $n$ distinct such cycles $\mathfrak{C}_a^{\prime}$. Combining them with the $n+1$ cycles $\mathfrak{C}_a$, we can thus build a basis consisting of $2n+1$ non-contractible four-cycles in the solutions defined by \eq{H-pairwise}.

Having identified a basis for non-contractible four-cycles, we are now in a position to derive the M-brane charges by integrating the four-form flux over the $\mathfrak{C}_a$. To ease the computation, we abstractly write the flux as \cite{Faedo:2020nol}
\begin{equation}\label{eq:F4-abs}
\begin{split}
    \mathcal{F}_{(4)} &= 2h_1\operatorname{vol}_{\text{AdS}_3}\wedge \text{d}z + 2h_1 \operatorname{vol}_{\mathds{S}^3}\wedge \text{d}z\\
    & + \partial_z H\rho^3 \text{d}\rho\wedge \operatorname{vol}_{\tilde{\mathds{S}}^3}-\partial_\rho H\rho^3 \text{d}z\wedge \operatorname{vol}_{\tilde{\mathds{S}}^3}~.
\end{split}
\end{equation}
The pull-back of the four-form field strength $\mathcal{F}_{(4)}$ onto any of the $\mathfrak{C}_a$ cycles eliminates the first two terms in \eq{F4-abs}. 
 Integrating along a given $\mathfrak{C}_a$ then yields
\begin{equation}
    \int_{\mathfrak{C}_a}P_{\mathfrak{C}_a}[\mathcal{F}_{(4)}]=2h_1\operatorname{Vol}({\tilde{\mathds{S}}^3})(\hat{\xi}_{2a}-\hat{\xi}_{2a-1})\,.
\end{equation}
The choice of orientation we made while defining the cycles $\mathfrak{C}_a$ ensures that the integral above is positive. This enables us to define the following charge
\begin{align}
    M_a &= \frac{1}{2(4\pi^2G_N)^{1/3}}\int_{\mathfrak{C}_a}P_{\mathfrak{C}_a}[\mathcal{F}_{(4)}]\,,
\end{align}
which can be interpreted as the number of M5-branes in the $a^{\rm th}$ stack \cite{Bachas:2013vza}, which obey $\sum_a M_a = M$ with $M$ being the total number of M5-branes.

It is also possible to generalize the construction above. As we mentioned previously, we can in fact build four-cycles surrounding more than one collapse point $\nu_a$. The four-cycle defined by $\mathfrak{C}_{bc} \equiv \sum_{a = b}^{c} \mathfrak{C}_a$, with $1\leq b\leq c\leq n+1$, is also non-contractible by construction, and is characterized by a charge
\begin{equation}
    \int_{\mathfrak{C}_{bc}}P_{ \mathfrak{C}_{bc}}[\mathcal{F}_{(4)}]=2h_1\operatorname{Vol}({\tilde{\mathds{S}}^3})\sum_{a=b}^{c}(\hat{\xi}_{2a}-\hat{\xi}_{2a-1})\,,
\end{equation}
under the four-form field strength.

We can follow a similar analysis for the flux threading the other set of non-contractible four-cycles, which we labelled $\mathfrak{C}^\prime_a$ earlier. In this case, only the second term in \eq{F4-abs} provides a non-vanishing contribution after pulling back the four-form field strength $\mathcal{F}_{(4)}$ to the cycle $\mathfrak{C}^\prime_a$.  Integrating the flux through $\mathfrak{C}^\prime_a$ gives 
\begin{equation}\label{eq:C4p-definition}
    \int_{\mathfrak{C}^\prime_a}P_{\mathfrak{C}^\prime_a}[\mathcal{F}_{(4)}]=2h_1\operatorname{Vol}({\mathds{S}^3})(\nu_{2a+1}-\nu_{2a-1})~.
\end{equation}
Similar to the integrated fluxes through the $\mathfrak{C}_a$ cycles, we define the charge
\begin{align}
    M^\prime_a &= \frac{1}{2(4\pi^2G_N)^{1/3}}\int_{\mathfrak{C}^\prime_a}P_{\mathfrak{C}^\prime_a}[\mathcal{F}_{(4)}]\,,
\end{align}
which is read as the number of M5$^\prime$-branes in the $a^{\rm th}$ stack (see Table \ref{tab:braneintersection}).

Again, we can easily generalize this analysis to four-cycles that connect non-neighboring collapse points $\nu_a$.  Denoting the sum of four-cycles as $\mathfrak{C}^\prime_{bc}\equiv\sum_{a=b}^c \mathfrak{C}^\prime_a$, where given the construction of the $\mathfrak{C}_a^\prime$ in \eq{C4p-definition} $1\leq b\leq c\leq n$, the integral of the flux through this cycle is simply
\begin{equation}
    \int_{\mathfrak{C}^\prime}P_{ \mathfrak{C}^\prime}[\mathcal{F}_{(4)}]=2h_1\operatorname{Vol}({\mathds{S}^3})\sum_a(\nu_{2a+1}-\nu_{2a-1})\,.
\end{equation}

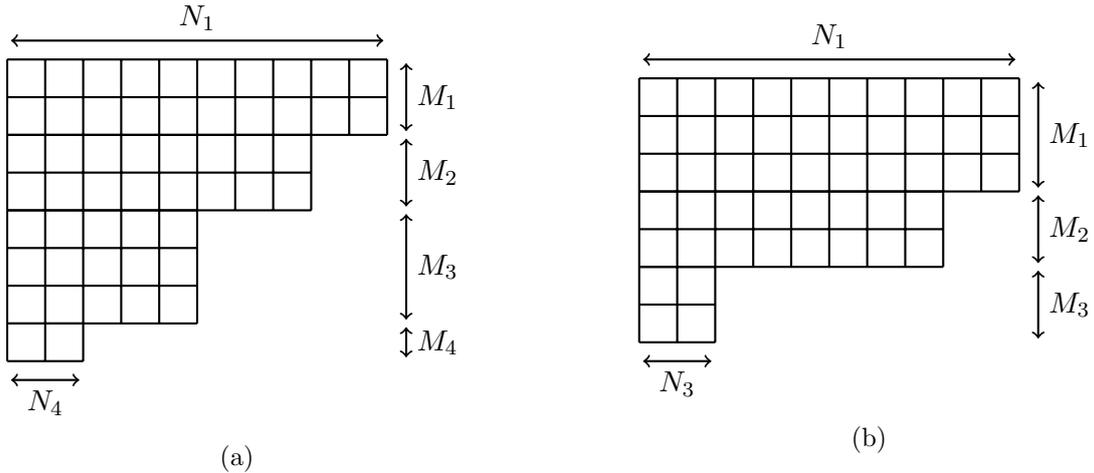
\begin{figure}[t]
     \centering
     \begin{subfigure}[h]{0.45\textwidth}
         \centering
         \begin{tikzpicture}[scale=0.5]
        \draw[thick,<->] (-4.9,5.5) -- (4.9,5.5) node[midway,above]{$N_1$};
        \draw[thick] (-5,5) grid (5,3);
        \draw[thick,<->] (5.5,4.9) -- (5.5,3.1) node[midway,right]{$M_1$};
        \draw[thick] (-5,3) grid (3,1);
        \draw[thick,<->] (5.5,2.9) -- (5.5,1.1) node[midway,right]{$M_2$};
        \draw[thick] (-5,1) grid (0,-2);
        \draw[thick,<->] (5.5,0.9) -- (5.5,-1.9) node[midway,right]{$M_3$};
        \draw[thick] (-5,-2) grid (-3,-3);
        \draw[thick,<->] (5.5,-2.1) -- (5.5,-2.9) node[midway,right]{$M_4$};
        \draw[thick,<->] (-4.9,-3.5) -- (-3.1,-3.5) node[midway,below]{$N_4$};

    \end{tikzpicture}
         \caption{}
         \label{fig:Young-Diagram-Pairwise}
     \end{subfigure}
     \hfill
     \begin{subfigure}[h]{0.45\textwidth}
         \centering
         \begin{tikzpicture}[scale=0.5]
        \draw[thick,<->] (-4.9,5.5) -- (4.9,5.5) node[midway,above]{$N_1$};
        \draw[thick] (-5,5) grid (5,2);
        \draw[thick,<->] (5.5,4.9) -- (5.5,2.1) node[midway,right]{$M_1$};
        \draw[thick] (-5,2) grid (3,0);
        \draw[thick,<->] (5.5,1.9) -- (5.5,0.1) node[midway,right]{$M_2$};
        \draw[thick] (-5,0) grid (-3,-2);
        \draw[thick,<->] (5.5,-0.1) -- (5.5,-1.9) node[midway,right]{$M_3$};
        \draw[thick,<->] (-4.9,-2.5) -- (-3.1,-2.5) node[midway,below]{$N_3$};
    \end{tikzpicture}
         \caption{}
         \label{fig:Young-Diagram-Multi}
     \end{subfigure}
        \caption{(a) Young diagram corresponding to the partition specifying a $\gamma\to-\infty$ solution with $5$ distinct $\nu_a$ constructed by pairwise collapse of 10 different $\xi_j$.  (b) Young diagram obtained by ``multiwise" collapse of 10 different $\xi_j$ to $4$ distinct $\nu_a$.  Another way to realize the construction of (b) starts from the partition in (a) and collapses $\nu_2=\nu_3$.
        }
        \label{fig:three graphs}
\end{figure}

In addition, following \cite{Bachas:2013vza} we can deduce the number of M2-branes ending on the $a^\text{th}$ stack of M5-branes, which we denote
\begin{equation}
    N_a=\sum_{b=a}^{n}M^\prime_b=\frac{h_1\operatorname{Vol}({\mathds{S}^3})}{(4\pi^2G_N)^{1/3}}\sum_{b=a}^n(\nu_{2b+1}-\nu_{2b-1})\,.
\end{equation}
One may notice how the definition of the collapse points $\nu_a$ influences the properties of $N_a$. Indeed, since $\{\nu_a\}$ is an ordered set, we see that $N_a\geq N_b$ for $a\leq b$. In other words, the set $\{N_a\}$ forms a partition of the total number of M2-branes $N= \sum_a N_a$, and so it is possible to define a Young diagram to encode the brane charges as illustrated in \fig{Young-Diagram-Pairwise}.

We can now recast the various moments defined in eqs.~\ref{eq:mhat-moments} and \ref{eq:moments-n} in terms of this partition data. Since only $\hat{m}_1$ appears in the asymptotic expansions in \app{FG}, and so also in the physical quantities computed using those expansions, we will not treat the higher $\hat{m}_j$ moments and simply write
\begin{equation}\label{eq:moment-m-branes}
    \hat{m}_1=\sum_{j=1}^{2n+2}(-1)^j\hat{\xi}_j=\frac{(4\pi^2G_N)^{1/3}}{h_1\operatorname{Vol}({\tilde{\mathds{S}}^3})}M\,,
\end{equation}
which gives back the usual relation between $M$ and the length scale on the $\mathds{S}^4$,
\begin{equation}
   L^3_{\mathds{S}^4} = \frac{h_1\hat{m}_1}{2}=\frac{(G_N)^{1/3}}{(2\pi)^{4/3}}M\,.
\end{equation}
The $\hat{n}_j$ moments are a bit trickier to re-express in terms of $M_a$ and $N_a$, but one can show that the following relation holds for any $j$
   \begin{align}\label{eq:moment-n-branes}
   \sum_{k=0}^j(-1)^{j+k}\frac{j!}{k!(j-k)!} \nu_{2n+1}^k ~\hat{n}_{j-k} = \left(\frac{G_N^{1/3}}{h_1\pi(2\pi)^{1/3}}\right)^{j+1}~\sum_{a=1}^{n+1}M_aN_a^j~,
    \end{align}
where it is understood that $\hat{n}_0= \hat{m}_1$.  However, we will only require relations up to $j=2$ moving forward. Using the expressions above, we can write
\begin{align}\label{eq:moment-n-branes-2}
    \hat{n}_1^2- \hat{m}_1\hat{n}_2 = \frac{G_N^{4/3}}{h_1^4\pi^4(2\pi)^{4/3}}\left[\left(\sum_{a=1}^{n+1}M_aN_a\right)^2 - M \sum_{a=1}^{n+1}M_a N_a^2\right]~,
\end{align}
which will be useful in the following section.

 Before moving on, we recall from our previous discussion that it is also possible to consider solutions where more than two branch points $\xi_j$ collapse to a single point $\nu_a$ and, of course, build non-contractible four-cycles around or connecting them. Let us again index the $p$ distinct loci of collapse as $\nu_a$, where now $1\leq a\leq p+1$ with $p\leq n$. The limiting case $p=n$ recovers the pairwise collapse described above. It will be useful to label $I_a$ and $K_a$ respectively as the smallest and largest $j$-indices of the $\xi_j$ branch points which collapse to a given $\nu_a$, i.e. $\nu_a:=\nu_{I_a}=\nu_{I_a+1}=\cdots=\nu_{K_a}$. The ordering of the collapse points $\nu_a$ and of the branch points $\xi_j$ was discussed previously in \sn{newsmall}; in particular, we recall that the parameters $\hat{\xi}_j$'s associated to the branch points collapsing to the same $\nu_a$ are ordered amongst themselves. In the end, the analysis for these ‘‘multiwise'' collapse solutions is identical to the one presented above for the pairwise collapse scenario, up to the replacements
 \begin{align}
     \hat{\xi}_{2a}-\hat{\xi}_{2a-1}\quad\longrightarrow\quad \sum_{j=I_a}^{K_a}(-1)^j\hat{\xi}_j~
 \end{align}
 and $n\to p$ throughout the expressions above. The net effect is that the partition data yields a differently shaped Young diagram, illustrated in \fig{Young-Diagram-Multi}. In particular, since the multiwise collapse can greatly reduce the number of singular points on the boundary of the Riemann surface, we see that the Young tableaux specifying the $\gamma\to-\infty$ solutions have at most $p$ rows. This is a striking difference compared to the Young tableau construction of \cite{Estes:2018tnu} for the Wilson surface solutions with $\gamma=-1/2$, whose associated Young tableaux always have $n$ rows. This difference is maximal in the 1PV solution described in \sn{1PV}, to which one cannot attach any Young tableau interpretation at all. This is due to the fact that the 1PV solution features a single collapse point $\nu$, so that the construction of the $\mf{C}_a^\prime$ cycle fails, thus preventing the definition of the $M_a^\prime$ and $N_a$ charges.

%%%%%%%%%%%%%%%%%%%%%%%%%%%%%%%%%%%%%%%%%%%%%%%%%%
%%%%%%%%%%%%%%%%%%%%%%%%%%%%%%%%%%%%%%%%%%%%%%%%%%
\section{\texorpdfstring{\boldmath Entanglement entropy of small $\N=(4,4)$ surface defects}{Entanglement entropy of small N=(4,4) surface defects}}
\label{sec:EE}
%%%%%%%%%%%%%%%%%%%%%%%%%%%%%%%%%%%%%%%%%%%%%%%%%%
%%%%%%%%%%%%%%%%%%%%%%%%%%%%%%%%%%%%%%%%%%%%%%%%%%

The entanglement entropy $S_{\text{EE}}$ of a spatial subregion $\mathcal{B}$ within a QFT is defined as the von Neumann entropy of the reduced density matrix obtained by tracing out the states in the complementary region $\overline{\mathcal{B}}$ of the QFT. For CFTs with weakly coupled gravity duals, the RT prescription \cite{Ryu:2006bv,Ryu:2006ef,Nishioka:2021uef} holographically recasts the computation of $S_{\text{EE}}$ into the following Plateau problem in the asymptotically AdS bulk,
\begin{equation}
    S_{\text{EE}}=\min_{\zeta}\frac{\mathcal{A}[\zeta]}{4G_{\text{N}}}~,
\end{equation}
where $\mathcal{A}[\zeta]$ is the area functional evaluated on a bulk hypersurface $\zeta$ which is homologous to the chosen spatial subregion in the dual CFT: $\zeta\cup\mathcal{B}=\partial b$ for some static bulk subregion $b$. We will denote this extremal bulk hypersurface as $\zeta_{\text{RT}}$. In the computations below, we choose the spatial subregion to be an Euclidean 5-ball, $\mathcal{B}=\mathbb{B}_R^5\hookrightarrow\mathbb{R}^5$, with radius $R$. We take $\mathcal{B}$ to be centered on the spatial extent of the surface defect, which has a Lorentzian worldvolume $\Upsilon_2=\mathbb{R}^{1,1}$.

In an ordinary QFT, the presence of highly entangled UV degrees of freedom induces short-distance divergences near the surface $\partial\mathcal{B} = \mathds{S}^4$.  Most of the divergences in the EE of a general QFT are non-universal and shape-dependent. However, in even dimensional theories, there are universal log-divergent contributions to the EE, which are generically related at conformal fixed points to the Weyl anomalies of the CFT.   

This is true in the presence of a defect as well, but we see additional divergences that arise from the defect degrees of freedom near $\Upsilon_2\cap \partial\mathcal{B}$.  In order to isolate these defect-localized contributions, we will adopt a scheme where we subtract off the EE of the deformed, vacuum ambient CFT, $S_{\rm EE}[\emptyset]$, from the EE computed in the presence of the defect $S_{\rm EE}[\Upsilon_2]$.  We can then extract the coefficient of the universal, log-divergent part of the defect contribution to the sphere EE by 
\begin{align}\label{eq:defect-EE-universal-formula}
    S_{\rm EE}^{\rm (univ)}&=R\frac{\text{d}}{\text{d}R}\left(S_{\text{EE}}[\Upsilon_2]-S_{\text{EE}}[\emptyset]\right)\big|_{R\to 0}~.
\end{align}
 In \cite{Jensen:2018rxu}, it was shown that contribution to the log-divergent part of the EE of a spherical region coming from a flat, $2$d conformal defect embedded in a $d$-dimensional flat-space ambient CFT takes the form of a linear combination of defect localized Weyl anomalies.  Explicitly, for a $6$d ambient CFT
\begin{align}\label{eq:defect-ee-anomalies}
    S_{\rm EE}^{\rm(univ)} = \frac{1}{3}\left(a_\Upsilon - \frac{3}{5}d_2\right)~.
\end{align}
where $a_{\Upsilon}$ is the A-type defect Weyl anomaly coefficient--i.e. appearing with the intrinsic Euler density--and $d_2$ is the B-type anomaly that enters with the trace of the pullback of the ambient Weyl tensor.  Importantly, while \cite{Jensen:2018rxu} demonstrated that $d_2\geq0$ based on energy conditions, $a_\Upsilon$, though obeying a defect ``c-theorem'', has no positivity constraints.   This means that as opposed to an ordinary, unitary $2$d CFT where the universal part of the EE is is proportional to the central charge \cite{Holzhey:1994we} and, hence, is non-negative, it is clear from \eq{defect-ee-anomalies} that $S_{\rm EE}^{\rm(univ)}$ is not similarly bounded nor is it RG monotonic.

 In completing the holographic computation, we also need to contend with the fact that the FG expansion is not globally defined as it typically breaks down in a region near the AdS submanifold that is dual to the insertion of the defect in the field theory.  However, we have a full analysis of the asymptotic expansions of the data specifying the $\gamma\to-\infty$ solutions in \app{FG}, and we have the general prescription for the FG transformation suitable for defect EE in \cite{Estes:2014hka}.  Together, we will be able to unambiguously holographically compute $S_{\rm EE}^{\rm (univ)}$.

To begin the holographic computation of the defect EE, we choose the following parametrization for the AdS$_3$ subspace in \eq{FG-metric},
\begin{equation}
\text{d}s^2_{\text{AdS}_3}=\frac{1}{u^2}\left(\text{d}u^2-\text{d}t^2+\text{d}x^2_\parallel\right).
\end{equation}
Furthermore, we take the RT hypersurface $\zeta$ to wrap both $\mathds{S}^3$ and $\tilde{\mathds{S}}^3$, and its profile in the remaining subspace to be described by $x_\parallel(u,\rho,z)$. The area of $\zeta$ as measured against the metric in \eq{FG-metric} is then 
\begin{equation}
	A[\zeta]=\operatorname{Vol}(\mathds{S}^3)\operatorname{Vol}(\tilde{\mathds{S}}^3)\int \text{d}u\int_{\Sigma_2}\ \text{d}\rho \text{d}z\ \mathcal{L}~,
\end{equation}
where the Lagrangian is
\begin{align}
\mathcal{L}&=\frac{h_1^2\rho^3}{u^2}\left[h_1\left((\partial_\rho x_\parallel)^2+(\partial_z x_\parallel)^2\right)H(z,\rho)+ u^2\left(1+(\partial_ux_\parallel)^2\right)H(z,\rho)^2\right]^{1/2}.
\end{align}
As shown in \cite{Jensen:2013lxa}, the minimal area surface wraps the Riemann surface $\Sigma_2$ too, so that $\partial_\rho x_\parallel=\partial_z x_\parallel=0$. The Lagrangian is thus minimized by
\begin{equation}
	x_\parallel^2+u^2=R^2,
\end{equation}
for a constant $R$, so that the area of the extremal hypersurface $\zeta_{\text{RT}}$ is
\begin{equation}\label{eq:RT-Integral}
	A[\zeta_{\text{RT}}]=h_1^3\operatorname{Vol}(\mathds{S}^3)\operatorname{Vol}(\tilde{\mathds{S}}^3)\log\left(\frac{2R}{\epsilon_u}\right)\int_{\Sigma_2}\text{d}\rho\text{d}z \ \rho^3\sum_{j=1}^{2n+2}\frac{(-1)^j\hat{\xi}_j}{\left(\rho^2+(z-\nu_j)^2\right)^{3/2}} +\mathcal{O}(\epsilon_u^2),
\end{equation}
where we introduced a small-$u$ cut-off $\epsilon_u>0$.

The evaluation of the integral above is performed in detail in \app{HEE-integral}. The resulting holographic entanglement entropy is
\begin{equation}\label{eq:RT-Area}
    S_{\text{EE}}[\Upsilon_2]=
    \frac{\pi^4L_{\mathds{S}^4}^9}{G_\text{N}}\log\left(\frac{2R}{\epsilon_u}\right)\left[\frac{64}{3}\frac{1}{\epsilon_v^4}+\frac{16}{5}\frac{\hat{n}_1^2}{\hat{m}_1^4}-\frac{16}{5}\frac{\hat{n}_2}{\hat{m}_1^3}+\mathcal{O}(\epsilon_v^2)\right]+\mathcal{O}(\epsilon_u^2),
\end{equation}
where $\epsilon_v>0$ is a small-$v$ cutoff in the FG parametrization.

Subtracting off the 1PV contribution to the entanglement entropy, $S_{\text{EE}}^{\text{1PV}}$, precisely removes the $\epsilon_v^{-4}$ divergence from \eq{RT-Area}, and leaves the $\mathcal{O}(\epsilon_v^0)$ term unchanged.  To see this, we recall that the 1PV limit takes $\hat{n}_k\to 0$, which in \eq{RT-Area} gives
\begin{align}\label{eq:RT-Area-1PV}
    S_{\text{EE}}^{\rm 1PV} =  \frac{\pi^4L_{\mathds{S}^4}^9}{G_\text{N}}\log\left(\frac{2R}{\epsilon_u}\right)\left[\frac{64}{3}\frac{1}{\epsilon_v^4}+\mathcal{O}(\epsilon_v^2)\right]+\mathcal{O}(\epsilon_u^2).
\end{align}
Therefore, we can at once compute the coefficient of the universal part of the defect sphere EE  using \eq{defect-EE-universal-formula} and plugging in eqs.~(\ref{eq:RT-Area}) and (\ref{eq:RT-Area-1PV}) with $S_{\rm EE}[\emptyset]= S_{\rm{EE}}^{\rm 1PV}$ to find
\begin{subequations}\label{eq:SEE-universal-defect}
\begin{align}
S_{\rm EE}^{\rm(univ)}  &= \frac{16}{5}\frac{\pi^4 L_{\mathds{S}^4}^9}{ G_{\text{N}}}\frac{\hat{n}_1^2-\hat{m}_1\hat{n}_2}{\hat{m}_1^4}\\
&=\frac{ 1}{5M}\left[\left(\sum_{a=1}^{n+1}M_aN_a\right)^2-M\sum_{a=1}^{n+1}M_aN_a^2\right],
\end{align}
\end{subequations}
where we have mapped to field theory quantities using $L_{\mathds{S}^4}^3 =\frac{G_N^{1/3}}{(2\pi)^{4/3}}M $ and used the definitions of moments $\hat{n}_j$, $\hat{m}_j$ in terms of the numbers of branes in eqs.~(\ref{eq:moment-m-branes}) and (\ref{eq:moment-n-branes-2}). In terms of the highest weight $\varpi$ of the $A_{M-1}$ irreducible representation encoded in the Young diagrams that specify the defect discussed in the previous section, we can re-express the defect sphere EE as \cite{Estes:2018tnu}
\begin{align}
    S_{\rm EE}^{\rm(univ)} = -\frac{(\varpi,\varpi)}{5}~,
\end{align}
where $(\cdot,\cdot)$ is the scalar product on the weight space induced by the Killing form.

We also note that the contribution of the aforementioned bulk deformation to the coefficient of the universal, log-divergent component of the vacuum-subtracted entanglement entropy is
\begin{align}
    S_{\rm EE}^{\rm(univ,bulk-def.)}  &= \frac{8}{3}\frac{\pi^4 L_{\mathds{S}^4}^9}{ G_{\text{N}}}~ = \frac{M^3}{6}~.
\end{align}
This is independent of the moments $(\hat{m}_i,\hat{n}_j)$, in line with the lack of Young Tableau data associated to the bulk deformation. Had we subtracted in \eq{defect-EE-universal-formula} the AdS$_7\times\mathds{S}^4$ vacuum, rather than the 1PV, the resulting entanglement entropy would have received both the defect and bulk deformation contributions above.

Finally, we note that the same quantity can be trivially computed in the orbifolded theory described in \sn{orbifolds} simply by rescaling 
\begin{eqnarray}
    S_{\rm EE}^{\rm(univ)}\longrightarrow\frac{\operatorname{Vol}(\mathds{S}^3/\mathbb{Z}_k)\operatorname{Vol}(\tilde{\mathds{S}}^3/\mathbb{Z}_{k^\prime})}{\operatorname{Vol}(\mathds{S}^3)\operatorname{Vol}(\tilde{\mathds{S}}^3)}S_{\rm EE}^{\rm(univ)}=\frac{S_{\rm EE}^{\rm(univ)}}{kk^\prime}~.
\end{eqnarray}

%%%%%%%%%%%%%%%%%%%%%%%%%%%%%%%%%%%%%%%%%%%%%%%%%%
%%%%%%%%%%%%%%%%%%%%%%%%%%%%%%%%%%%%%%%%%%%%%%%%%%
\section{Summary and Outlook}
\label{sec:Discussion}

In this work, we have constructed a novel class of solutions in 11d SUGRA that are holographically dual to 2d superconformal defects preserving small $\mathcal{N}=(4,4)$ and $\mathcal{N}=(0,4)$ SUSY in 6d SCFTs at large $M$.  These solutions fit into the one-parameter family organized in a general classification scheme of 11d SUGRA solutions with superisometry $\mathfrak{d}(2,1;\gamma)\oplus\mathfrak{d}(2,1;\gamma)$ \cite{Bachas:2013vza}; specifically, they are obtained in the $\gamma\to-\infty$ limit.  There are several features of these new solutions that separate them from the more familiar $\gamma=-1/2$ case that holographically corresponds to $1/2$-BPS Wilson surface type defects in the 6d $\mathcal{N}=(2,0)$ $A_{M-1}$ SCFT.  

Within the one-parameter family of solutions labelled by $\gamma$, the $\gamma\to-\infty$ limit is slightly unusual from the superalgebra perspective.  Despite producing an AdS$_7\times\mathds{S}^4$ asymptotic geometry as shown in \sn{large2small}, taking the $\gamma\to-\infty$ limit means that $\mathfrak{d}(2,1;\gamma)\oplus\mathfrak{d}(2,1;\gamma)$ is not a subalgebra of the $\mathfrak{osp}(8^*|4)$ superisometry of AdS$_7\times\mathds{S}^4$.  On the field theory side of the holographic duality, this means that the ambient theory into which the defects are inserted is some  deformation of the 6d $\N=(2,0)$ SCFT.  

We have seen that choosing all of the singular loci in the internal space to collapse to a single point -- a configuration which we label 1PV -- destroys the data that specifies the defect, i.e. the Young diagram corresponding to the arrangement of M5-branes. However, as is clear from the discussion in \sn{1PV}, the vacuum that we arrive at has an isometry group of $SO(2,2)\times SO(3)\times SO(5)$, as opposed to the vacuum solution at $\gamma=-1/2$, which instead enjoys the full $SO(6,2)$.  In the latter case, the trivial defect corresponds to a Wilson surface transforming in the $\bf{1}$ of $A_{M-1}$, and the conformal symmetry of the ambient 6d $\N=(2,0)$ theory is restored.  For $\gamma\to-\infty$, the trivial defect dual to the 1PV still possesses what looks like the `defect' conformal symmetry despite being the vacuum solution, which renders giving a precise definition for and interpretation of the defect CFT difficult. The 1PV does, however, enable us to correctly employ a background subtraction scheme\footnote{Here, ``correctly'' refers to a background subtraction which also removes any contributions from the trivial defect. As we have demonstrated, the same cannot be said of a subtraction scheme which utilizes vacuum AdS$_7\times \mathds{S}^4$.} and arrive at a finite result for $S_{\rm EE}^{\rm (univ)}$ and, we believe, resolves the puzzling appearance of divergences in the ``defect central charge'' computed in \cite{Faedo:2020nol}.

On a more fundamental level, the small $\N=(4,4)$ defects at $\gamma\to-\infty$ cannot be viewed as a smooth deformation of the Wilson surface defects at $\gamma=-1/2$. Indeed, the $\gamma\to-\infty$ solutions cannot even be smoothly deformed into the solution at $\gamma=0$, to which they are related by the involution $\gamma\mapsto1/\gamma$ with an exchange of the $\mathfrak{so}(3)\oplus\mathfrak{so}(3)$ factors in $\mathfrak{d}(2,1;\gamma;0)\oplus\mathfrak{d}(2,1;\gamma;0)$.  The reason is that there is a special point at $\gamma= -1$ where the real form $\mathfrak{d}(2,1;\gamma;0)$ becomes $\mathfrak{osp}(4|2;\mathbb{R})$. At this value of $\gamma$, $SO(2,2)$ Wigner-\.{I}n\"{o}n\"{u} contracts to $ISO(1,2)$, and AdS$_3$ becomes $\mathbb{R}^{2,1}$.  Therefore, the $\gamma\to-\infty$ solutions are isolated from the other class of asymptotically AdS$_7\times\mathds{S}^4$ geometries.

In light of the new small $\mathcal{N}=(4,4)$ solutions that we have constructed and holographically studied, there are a number of open questions that remain to be answered. 

Firstly, as we discussed at the start of \sn{EE} , the contribution from a flat 2d conformal defect to the log-divergent, universal part of the EE of a spherical region in a $d\geq 4$ ambient CFT is built from a linear combination of two defect Weyl anomaly coefficients, $a_\Upsilon$ and $d_2$ that characterize the defect theory.  In order to disentangle these two fundamental defect quantities, we would compute a second holographic quantity that contains either $a_\Upsilon$ or $d_2$.  For instance, $d_2$ controls the normalization of the one-point function $\left<T_{\mu\nu}\right>$ of the stress-energy tensor, which  can be readily computed for most 10d or 11d supergravity solutions by dimensional reduction on the internal space \cite{deHaro:2000vlm, Skenderis:2006uy}. Therefore, it is natural to try to compute $d_2$ and $S_{\rm EE}^{\rm (univ)}$ in order to isolate the independent defect Weyl anomalies\footnote{In fact, $a_{\Upsilon}$ and $d_2$ are the only independent defect Weyl anomaly coefficients for superconformal defects preserving at least 2d $\mathcal{N}=(0,2)$ supersymmetry.  This was shown for co-dimension four defects in 6d SCFTs in \cite{Drukker:2020atp} and co-dimension two defects in 4d SCFT in \cite{Bianchi:2019sxz}. }. This was successfully done for the Wilson surfaces at $\gamma=-1/2$ in \cite{Estes:2018tnu} and for codimension-2 defects in \cite{Capuozzo:2023fll}. However, the $\gamma\to-\infty$ solutions are more subtle, and a na\"ive application of dimensional reduction techniques would be inappropriate.  Namely, in reducing the 11d solutions to 7d, the presence of the non-trivial four-form flux modifies the gravitational equations of motion at the conformal boundary of AdS$_7$, which violates the assumptions in \cite{deHaro:2000vlm}. Therefore, computing $d_2$ holographically from $\left<T_{\mu\nu}\right>$ requires a generalization to account for flux contributions, which is the subject of ongoing work.

Furthermore, it is natural to look for physical observables which can be reliably computed on the field theory side and employed to test the holographic predictions made above.  For 2d BPS conformal defects in 6d $A_{M-1}$ and $D_M$ $\N=(2,0)$ SCFTs at large $M$, recent developments in analytic bootstrap methods have enabled the computation of correlations functions in the presence of 2d defects that are controlled by anomalies \cite{Drukker:2020atp,Meneghelli:2022gps}. Further, despite the lack of a Lagrangian description and of supersymmetric localization methods for 6d SCFTs at large $M$, chiral algebra methods have also been shown to give exact results for defect correlators \cite{Meneghelli:2022gps} and the defect SUSY Casimir energy \cite{Bullimore:2014upa, Chalabi:2020iie}.  Currently, only Wilson surface type defects, i.e. the holographically dual theories to the $\gamma=-1/2$ solutions, have been studied using these field theory techniques. It is reasonable to wonder whether any of these methods are applicable to the types of defects in the deformed 6d theory that we have constructed in this work.

The biggest hurdle to clear in trying to generalize bootstrap or chiral algebra methods for use in the dual to the 1PV of the $\gamma\to-\infty$ solutions is clarifying the precise role of the deformation parameter $\gamma$.  As we have explained in the $\gamma\to-\infty$ limit, the ambient 6d theory has reduced global and conformal symmetries, and there is no smooth path in field theory space as $\gamma$ is varied through $\gamma=-1$ to get from the 6d $A_{M-1}$ $\N = (2,0)$ theory to the dual of the 1PV.  It is unclear at the moment precisely what symmetry breaking operators are sourced on the field theory side in the $\gamma\to-\infty$ limit.

%%%%%%%%%%%%%%%%%%%%%%%%%%%%%%%%%%%%%%%%%%%%%%%%%%
%%%%%%%%%%%%%%%%%%%%%%%%%%%%%%%%%%%%%%%%%%%%%%%%%%

\section*{Acknowledgments}
The authors would like to thank Yolanda Lozano and Nicol\`{o} Petri for useful discussions during the completion of this work and for their comments on a draft of this manuscript. The authors would also like to thank Andy O'Bannon for his contributions during the early phase of this research.
The work of PC is supported by a Mayflower studentship from the University of Southampton. The work of BR is supported by the INFN. The work of BS is supported in part by the STFC consolidated grant ST/T000775/1. This material is based upon work supported by the U.S. Department of Energy, Office of Science, Office of High Energy Physics under Award Number DE-SC0024557.

Disclaimer:  ``This report was prepared as an account of work sponsored by an agency of the United States Government.  Neither the United States Government nor any agency thereof, nor any of their employees, makes any warranty, express or implied, or assumes any legal liability or responsibility for the accuracy, completeness, or usefulness of any information, apparatus, product, or process disclosed, or represents that its use would not infringe privately owned rights.  Reference herein to any specific commercial product, process, or service by trade name, trademark, manufacturer, or otherwise does not necessarily constitute or imply its endorsement, recommendation, or favoring by the United States Government or any agency thereof.  The views and opinions of authors expressed herein do not necessarily state or reflect those of the United States Government or any agency thereof.''

\appendix
\section{Fefferman-Graham parametrization}\label{app:FG}

In this Appendix, we will be concerned with finding a reparametrization $\{w,\bar{w}\}\to\{v,\phi\}$ of the Riemann surface $\Sigma_2$ under which the 11d SUGRA solution in \eq{Johns-Solutions} can be asymptotically (in particular, for small $v$) recast into the following form,
\begin{equation}\label{eq:AlAdS}
	\text{d} s^2=\frac{4L_{\mathds{S}^4}^2}{v^2}\left[\text{d} v^2+\alpha_1\text{d} s^2_{\text{AdS}_3}+\alpha_2\text{d} s^2_{\mathds{S}^3}\right]+L_{\mathds{S}^4}^2\left[\alpha_3\text{d}\phi^2+\alpha_4s^2_{\phi}\text{d} s^2_{\tilde{\mathds{S}}^3}\right]~,
\end{equation}
for some $\mathcal{O}(v^0)$ metric factors $\{\alpha_i\}_{i=1}^4$. By imposing that the metric factor in the $v$ direction be exact in $v$, as above, we can reconstruct the asymptotic reparametrization order by order in $v$ and in terms of polar coordinates 
\begin{align}\label{eq:Polar-Coordinates-Sigma-2}
    r=\sqrt{z^2+\rho^2}\quad \text{and}\quad \theta=\arctan(\rho/z)
\end{align}
on $\Sigma_2$ to be the following,
\begin{subequations}\allowdisplaybreaks\label{eq:FG-map}
    \begin{align}\allowdisplaybreaks
        r(v,\phi)&=-\frac{2 (\gamma +1)^2 m_1}{\gamma  v^2}+\frac{(2 \gamma +1) m_1 \left(c_{2 \phi }-3\right)}{24 \gamma }+\frac{m_2 c_{\phi }}{2m_1}+\Biggl[\frac{3 \gamma ^2 m_2^2 \left(7 c_{2 \phi }+1\right)}{4 m_1^3}\\\nonumber
        &\qquad-\frac{\gamma ^2 m_3 \left(5 c_{2 \phi }+3\right)}{m_1^2}-\frac{2 \gamma  (2 \gamma
        +1) m_2 c_{\phi } s_{\phi }^2}{m_1}-\frac{1}{16} (8 \gamma  (\gamma +1)+3) m_1 c_{2 \phi }\\\nonumber
        &\qquad+\frac{37}{48} \gamma  (\gamma +1) m_1+\frac{73 m_1}{192}+\frac{19}{192} (2 \gamma +1)^2 m_1 c_{4 \phi }\Biggr]\frac{v^2}{48\gamma(\gamma+1)^2}+\mathcal{O}(v^4),\\
       \theta(v,\phi)&=\phi +\frac{(2 \gamma +1) m_1^2 c_\phi+3 \gamma m_2}{12 (\gamma +1)^2 m_1^2}s_\phi v^2 +\Biggl[\frac{9 \gamma ^2 m_2^2 s_{2 \phi }}{8 m_1^4}-\frac{5 \gamma ^2 m_3 s_{2 \phi }}{6 m_1^3}\\\nonumber
       &\qquad+\frac{c_{\phi } s_{\phi } \left(5 (2 \gamma +1)^2 c_{2 \phi }-24 \gamma  (\gamma
       +1)-7\right)}{48}\\\nonumber
       &\qquad+\frac{\gamma  (2 \gamma +1) m_2 \left(3 c_{2 \phi }-1\right) s_{\phi }}{12m_1^2}\Biggr]\frac{v^4}{16(\gamma+1)^4}+\mathcal{O}(v^6)~,
    \end{align}
\end{subequations}
where we introduced the moments
\begin{equation}
    m_k:=\sum_{j=1}^{2n+2}(-1)^j\xi_j^k~.
\end{equation}
Under this mapping, the metric takes the desired form of \eq{AlAdS} with the following metric factors
\begin{subequations}\allowdisplaybreaks\label{eq:alphaGeneralGamma}
    \begin{align}\allowdisplaybreaks
        \alpha_1(\gamma)&=1+\frac{2\gamma+3-(1+2\gamma)c_{2\phi}}{16
        (\gamma +1)^2}v^2+\left[9 \left(4 (3 \gamma -13) \gamma +16 \gamma ^2\kappa-17\right)\right.\\\nonumber
        &\quad\left.-67 (2 \gamma +1)^2 c_{4\phi}+12 \left(8 (3 \gamma +1) \gamma +20 \gamma ^2 \kappa+3\right) c_{2\phi}\right]\frac{v^4}{18432(\gamma+1)^4}+\mathcal{O}(v^6)~,\\
        \alpha_2(\gamma)&=\frac{(\gamma +1)^2}{\gamma ^2}+\frac{2\gamma-1-(1+2\gamma)c_{2\phi}}{16 \gamma ^2}v^2+\left[9 \left(4 (3 \gamma +19) \gamma +16 \gamma ^2 \kappa+47\right)\right.\\\nonumber
        &\quad\left.-67 (2 \gamma +1)^2 c_{4\phi}+12 \left(8 (3 \gamma +5) \gamma +20 \gamma ^2\kappa+19\right) c_{2\phi}\right]\frac{v^4}{18432\gamma^2(\gamma +1)^2}+\mathcal{O}(v^6)~,\\
        \alpha_3(\gamma)&=1+\frac{(2 \gamma +1) c_{\phi}^2}{4 (\gamma +1)^2}v^2+\left[-12 (\gamma +1) \gamma -24 \gamma ^2\kappa-9+6 (2 \gamma +1)^2 c_{2\phi}\right.\\\nonumber
        &\quad\left.+7 (2 \gamma +1)^2 c_{4\phi}\right]\frac{v^4}{768(\gamma +1)^4}+\mathcal{O}(v^6)~,\\
        \alpha_4(\gamma)&=1+\frac{(2 \gamma +1)(2 c_{2\phi}+1)}{12 (\gamma +1)^2}v^2+\left[-52 (\gamma +1) \gamma -24 \gamma ^2\kappa-19\right.\\\nonumber
        &\quad\left.+79 (2 \gamma +1)^2 c_{4\phi}+12 \left(-2 (\gamma +1) \gamma -10 \gamma ^2\kappa-3\right) c_{2\phi}\right]\frac{v^4}{4608(\gamma +1)^4}+\mathcal{O}(v^6)~,
    \end{align}
\end{subequations}
where, to the order shown, the moments $\{m_i\}$ enter the metric factors only in the combination
\begin{equation}
    \kappa\equiv\frac{3 m_2^2-4 m_1 m_3}{m_1^4}~.
\end{equation}
Note that $\kappa$ is invariant under coordinate transformations $\{\xi_i\mapsto\xi_i+\lambda\}$, even though $m_2$ and $m_3$ are not individually. From this mapping, we also deduce the curvature scale of the asymptotic local $\mathds{S}^4$ in FG gauge,
\begin{equation}
    L_{\mathds{S}^4}^3=\frac{|1+\gamma|^3}{\gamma^{2}}\frac{h_1m_1}{2}~.
\end{equation}

We can now take the $\gamma\to-\infty$ limit, accompanied by appropriate rescalings as described in \sn{newsmall}, to determine the asymptotic local behavior of the metric in \eq{metric}. In this limit, we find that
\begin{equation}
    \frac{\kappa}{\gamma^2}\to\frac{12(\hat{n}_1^2-\hat{m}_1\hat{n}_2)}{\hat{m}_1^4}~,
\end{equation}
where we introduced the additional moments
\begin{equation}\label{eq:moments-n}
    \hat{n}_k:=\sum_{j=1}^{2n+2}(-1)^j\nu_j^k\hat\xi_j~.
\end{equation}
Furthermore, we find that this limiting procedure trivializes the relative warping between AdS$_3$ and $\mathds{S}^3$; that is, $\alpha_1(\gamma\to-\infty)=\alpha_2(\gamma\to-\infty)$. Overall, the limit $\gamma\to-\infty$ produces the FG line element advertized in \eq{FG-metric}, with the following metric factors,
\begin{subequations}\allowdisplaybreaks\label{eq:alphas}
    \begin{align}
        \alpha_1(\gamma\to-\infty)&=1+\frac{(3+5c_{2\phi})(\hat{n}_1^2-\hat{m}_1\hat{n}_2)}{32\hat{m}_1^4}v^4\\\nonumber&\quad-\frac{5(1+7c_{2\phi})(2\hat{n}_1^3-3\hat{m}_1\hat{n}_1\hat{n}_2+\hat{m}_1^2\hat{n}_3)c_{\phi}}{144\hat{m}_1^6}v^6+\mathcal{O}(v^8),\\
        \alpha_3(\gamma\to-\infty)&=1+\frac{3(\hat{m}_1\hat{n}_2-\hat{n}_1^2)}{8\hat{m}_1^4}v^4+\frac{5(2\hat{n}_1^3-3\hat{m}_1\hat{n}_1\hat{n}_2+\hat{m}_1^2\hat{n}_3)c_{\phi}}{12\hat{m}_1^6}v^6+\mathcal{O}(v^8),\\
        \alpha_4(\gamma\to-\infty)&=1+\frac{(1+5c_{2\phi})(\hat{m}_1\hat{n}_2-\hat{n}_1^2)}{16\hat{m}_1^4}v^4\\\nonumber&\quad+\frac{5(5c_{\phi}+7c_{3\phi})(2\hat{n}_1^3-3\hat{m}_1\hat{n}_1\hat{n}_2+\hat{m}_1^2\hat{n}_3)}{144\hat{m}_1^6}v^6+\mathcal{O}(v^8)~,
    \end{align}
\end{subequations}
and with $\mathds{S}^4$ radius given by \eq{radius}.

Finally, we note that, if all points are taken to collapse to the origin, $\nu_j=0$ for $j\in\{1,2,\ldots,2n+2\}$, the asymptotic reparametrization in \eq{FG-map} becomes exact,
\begin{subequations}
	\begin{align}
		r(v,\phi)&=\frac{2\hat{m}_1}{v^2}\\
		\theta(v,\phi)&=\phi~.
	\end{align}
\end{subequations}

\section{Area of the Ryu-Takayanagi hypersurface}\label{app:HEE-integral}

In this appendix, we fill in the technical details of the computation of the holographic entanglement entropy in \eq{RT-Area}. 

We begin by treating the integral in \eq{RT-Integral}.  In order to exploit the geometry of the internal space and facilitate an easier path to evaluating the remaining integral,  we first transform to polar coordinates $\{r,\,\theta\}$, introduced in \eq{Polar-Coordinates-Sigma-2}, with $\theta\in[0,\pi]$.  We can then expand the summands in \eq{RT-Integral} on the complete basis of $L^2(0,\pi)$ functions spanned by the Legendre polynomials, $P_k(c_{\theta})$.  The harmonic function $H$ can then be written in two equivalent representations
\begin{equation}\label{eq:H-Legendre-Reps}
	H(r,\theta)=
	\begin{cases}
        \frac{h_1}{2}\sum_{j=1}^{2n+2}(-1)^j\hat{\xi}_j\left|\nu_j\right|^{-3}\left(\sum_{k=0}^\infty r^k\nu_j^{-k}P_k(c_{\theta})\right)^3, \quad r\in [0,|\nu_j|),\vspace{0.2cm}\\ 
		\frac{h_1}{2}\sum_{j=1}^{2n+2}(-1)^j\hat{\xi}_jr^{-3}\left(\sum_{k=0}^\infty r^{-k}\nu_j^{k}P_k(c_{\theta})\right)^3	,\quad r\in (|\nu_j|,\Lambda_r(\epsilon_v,0)],
	\end{cases}
\end{equation}
which converge when integrated in $r$ over their respective domains. We have also introduced in the second line of \eq{H-Legendre-Reps} a cutoff scale $\Lambda_r$ at large $r$, which from the transformation to FG gauge in \eq{FG-map} can be expressed in a small $\epsilon_v$ expansion as
\begin{equation}\label{eq:RT-FG-cutoff}
\Lambda_r(\epsilon_v,\theta)=\frac{2\hat{m}_1}{\epsilon_v^2}+\frac{\hat{n}_1c_{\theta}}{\hat{m}_1}+\frac{(3+5c_{2\theta})\hat{m}_1\hat{n}_2-(5+3c_{2\theta})\hat{n}_1^2}{16\hat{m}_1^3}\ \epsilon_v^2+\mathcal{O}(\epsilon_v^4).
\end{equation}
Thus, the remaining integral in the area functional can be partitioned into two separate contributions
\begin{equation}\label{eq:RT-Area-I1-I2}
	A[\zeta_{\text{RT}}]=\frac{32\pi^4L_{\mathds{S}^4}^9}{\hat{m}_1^3}\log\left(\frac{2R}{\epsilon_u}\right)\sum_{j=1}^{2n+2}(-1)^j\hat\xi_j\left(\mathcal{I}_j^{(1)}+\mathcal{I}_j^{(2)}\right)+\mathcal{O}(\epsilon_u^2),
\end{equation}
where both integrals
\begin{subequations}\label{eq:I1-I2}
    \begin{align}
        \mathcal{I}_j^{(1)}&=\int_0^\pi\text{d}\theta\int_0^{\left|\nu_j\right|}\text{d} r\ r^4s^3_{\theta}\left(\frac{1}{\left|\nu_j\right|}\sum_{k=0}^\infty\left(\frac{r}{\nu_j}\right)^kP_k(c_{\theta})\right)^3,\\
        \mathcal{I}_j^{(2)}&=\int_0^\pi\text{d}\theta\int_{|\nu_j|}^{\Lambda_r(\epsilon_v,\theta)}\text{d} r\ r\ s^3_{\theta}\left(\sum_{k=0}^\infty\left(\frac{\nu_j}{r}\right)^kP_k(c_{\theta})\right)^3,
    \end{align}
\end{subequations}
are convergent. 

At first glance, \eq{I1-I2} may not seem to put us in a better position to evaluate the integral, but we can exploit the properties of $P_k(c_{\theta})$ over the interval $\theta\in[0,\pi]$. In particular, we can make use of the orthogonality relation of the triple-product of Legendre polynomials
\begin{equation}\label{eq:Legendre-Identity-1}
\int_0^\pi\text{d}\theta\ s_{\theta}\ P_{\ell_1}(c_{\theta})P_{\ell_2}(c_{\theta})P_{\ell_3}(c_{\theta})=2\begin{pmatrix}\ell_1 & \ell_2 & \ell_3 \\ 0 & 0 & 0\end{pmatrix}^2,
\end{equation}
where the right-hand side is the Wigner $3j$-symbol. Since this particular $3j$-symbol has vanishing magnetic quantum numbers, if $\ell\equiv\ell_1+\ell_2+\ell_3$ is even and together the $\ell_i$ satisfy the triangle inequality, then it can neatly be expressed as
\begin{equation}
\begin{pmatrix}\ell_1 & \ell_2 & \ell_3 \\ 0 & 0 & 0\end{pmatrix}=
\frac{(-1)^{\ell/2}(\ell/2)!}{\sqrt{(\ell+1)!}} \prod_{i=1}^3\frac{\sqrt{(\ell-2\ell_i)!}}{(\ell/2-\ell_i)!}~.
\end{equation}
Otherwise, if $\ell$ is not even or the triangle inequality is violated, the $3j$-symbol vanishes. Additionally, the following identity
\begin{equation}\label{eq:Legendre-Identity-2}
    P_{\ell_1} P_{\ell_2} = \sum_{\ell_3=\lvert \ell_1-\ell_2\rvert}^{\ell_1+\ell_2}\begin{pmatrix}\ell_1 & \ell_2 & \ell_3 \\ 0 & 0 & 0\end{pmatrix}^2(2\ell_3+1)P_{\ell_3}~,
\end{equation}
is particularly useful in the evaluation of the $\theta$-integrals in \eq{I1-I2}, where for brevity we denoted $P_\ell\equiv P_\ell(c_{\theta})$. Eventually, after applying both eqs.~(\ref{eq:Legendre-Identity-1}) and (\ref{eq:Legendre-Identity-2}), we find that the $\theta$-integrals in $\mathcal{I}^{(1)}_{j}$ can be handled with the help of
\begin{equation}\label{eq:LegendrePPPs3-W3j}
    \int_0^\pi\text{d}\theta\ s^3_{\theta}P_{\ell_1}P_{\ell_2}P_{\ell_3} = \frac{4}{3}\begin{pmatrix}\ell_1 & \ell_2 & \ell_3 \\ 0 & 0 & 0\end{pmatrix}^2-\frac{4}{3}\sum_{k=\lvert 2-\ell_1\rvert}^{2+\ell_1}(2k+1)\begin{pmatrix}2 & \ell_1 & k \\ 0 & 0 & 0\end{pmatrix}^2\begin{pmatrix}k & \ell_2 & \ell_3 \\ 0 & 0 & 0\end{pmatrix}^2.
\end{equation}

The integrals $\mathcal{I}_j^{(1)}$ and $\mathcal{I}_j^{(2)}$ may be further simplified by considering the convergence properties of the sum, which allow us to integrate each term in $r$ separately. Doing so, we rapidly see the usefulness of the previous $\theta$-integral formulae, which appear explicitly as
\begin{subequations}
\begin{align}
    \mathcal{I}_j^{(1)} =& \sum_{\ell_1,\ell_2,\ell_3}\frac{1}{5+\ell}\frac{\nu_j^{\ell+2}}{|\nu_j|^\ell}\int_0^\pi\text{d}\theta\ s^3_{\theta}P_{\ell_1}P_{\ell_2}P_{\ell_3},\label{eq:I1-I2-rintegrated-1}\\
    \mathcal{I}_j^{(2)} =& \sum_{\substack{\ell_1,\ell_2,\ell_3\\\ell\neq 2}}\frac{\nu_j^\ell}{2-\ell}\int_0^\pi\text{d}\theta\ s^3_{\theta}P_{\ell_1}P_{\ell_2}P_{\ell_3}\left[\Lambda_r^{2-\ell}(\epsilon_v,\theta)-\frac{\nu_j^2}{|\nu_j|^\ell}\right]\nonumber\\
    &+\sum_{\substack{\ell_1,\ell_2,\ell_3\\\ell=2}}\nu_j^2\int_0^\pi\text{d}\theta\ s^3_{\theta}P_{\ell_1}P_{\ell_2}P_{\ell_3}\ln\left(\frac{\Lambda_r(\epsilon_v,\theta)}{\lvert\nu_j\rvert}\right)\,,\label{eq:I1-I2-rintegrated-2}
\end{align}
\end{subequations}

where we have isolated the $\ell=2$ mode in $\mathcal{I}^{(2)}_j$ in order to handle the potential log divergence as a separate case.

Starting with the sum on the second line of \eq{I1-I2-rintegrated-2}, we first expand the integral in small $\epsilon_v$ using \eq{RT-FG-cutoff}.  Using the integral formulae for the Legendre polynomials above, we find that the leading $\ln\left(2\hat{m}_1/\epsilon_v^2\right)$ divergence contains no additional $\theta$-dependence, and so due to the constraint that $\ell=2$, is weighted with $P_1^2+P_2$, and vanishes upon integration.  Thus, we find that
\begin{align}
    \sum_{j=1}^{2n+2}(-1)^j\hat{\xi}_j\sum_{\substack{\ell_1,\ell_2,\ell_3\\\ell=2}}\nu_j^2\int_0^\pi\text{d}\theta\ s^3_{\theta}P_{\ell_1}P_{\ell_2}P_{\ell_3}\ln\left(\frac{\Lambda_r(\epsilon_v,\theta)}{\lvert\nu_j\rvert}\right)=\mathcal{O}(\epsilon_v^4)~.
\end{align}
Hence, the only meaningful contributions to $A[\zeta_{\text{RT}}]$ from $\mathcal{I}^{(2)}_j$ come from the first line in \eq{I1-I2-rintegrated-2}.

Moving on to the cutoff-dependent integrand on the first line of \eq{I1-I2-rintegrated-2}, we can again utilize the small $\epsilon_v$ expansion in \eq{RT-FG-cutoff}.  Since the leading divergence in $\Lambda_r(\epsilon_v,\theta)$ is $\mathcal{O}(1/\epsilon_v^2)$, we can neglect any integral for $\ell>2$ as it will vanish as $\mathcal{O}(\epsilon_v^2)$.  Truncating to the sum to $\ell<2$, expanding in small $\epsilon_v$, and evaluating the sum over $j$, we find that the total contribution to $A[\zeta_{\text{RT}}]$ from the first sum in \eq{I1-I2-rintegrated-2} is 
\begin{align}
    \sum_{j=1}^{2n+2}(-1)^j\hat{\xi}_j\sum_{\substack{\ell_1,\ell_2,\ell_3\\\ell\neq 2}}\frac{\nu_j^\ell}{2-\ell}\int_0^\pi\text{d}\theta\ s^3_{\theta}P_{\ell_1}P_{\ell_2}P_{\ell_3}\Lambda_r^{2-\ell}(\epsilon_v,\theta)=\frac{8\hat{m}_1^3}{3\epsilon_v^4}+\frac{2\hat{n}_1^2}{5\hat{m}_1}+\mathcal{O}(\epsilon_v^2).
\end{align}

Finally, we treat the remaining sums in \eq{I1-I2-rintegrated-1} and the second term on the first line of \eq{I1-I2-rintegrated-2} together. Firstly, we note that the integral at $\ell=2$ in $\mathcal{I}_j^{(1)}$ vanishes due the integrand being of the form $P_1^2+P_2$. Secondly, we make use of \eq{LegendrePPPs3-W3j} explicitly and observe that only the $\ell=0$ term contributes. That is, if we decompose the sum as partial sums in $\ell$,
\begin{align}
    \sum_{\substack{\ell_1,\ell_2,\ell_3\\\ell\neq 2}}\left(\frac{1}{5+\ell}-\frac{1}{2-\ell}\right)\int_0^\pi\text{d}\theta\ s^3_{\theta}P_{\ell_1}P_{\ell_2}P_{\ell_3} &= \sum_{\substack{a=0\\a\neq 2}}^{\infty}\left(\frac{1}{5+a}-\frac{1}{2-a}\right)\sum_{\substack{\ell_1,\ell_2,\ell_3\\\ell= a}}\int_0^\pi\text{d}\theta\ s^3_{\theta}P_{\ell_1}P_{\ell_2}P_{\ell_3}\nonumber\\
    &=-\frac{2}{5}-\frac{5}{6}\sum_{\substack{\ell_1,\ell_2,\ell_3\\\ell= 1}}\int_0^\pi\text{d}\theta\ s^3_{\theta}P_{\ell_1}P_{\ell_2}P_{\ell_3}+\ldots,
\end{align}
we find that all the partial sums with $\ell>0$ vanish and $-2/5$ is the exact result.

Putting all of the results above together and taking the sum over $j$, we find
\begin{align}
    \sum_{j=1}^{2n+2}(-1)^j\hat{\xi}_j(\mathcal{I}_j^{(1)}+\mathcal{I}_j^{(2)})=\frac{8\hat{m}_1^3}{3\epsilon_v^4}+\frac{2}{5}\frac{\hat{n}_1^2-\hat{n}_2\hat{m}_1}{\hat{m}_1}+\mathcal{O}(\epsilon_v^2).
\end{align}
Plugging in to \eq{RT-Area-I1-I2}, the unregulated area of the RT hypersurface is 
\begin{align}
    A[\zeta_{\text{RT}}]=
    \pi^4L_{\mathbb{S}^4}^9\log\left(\frac{2R}{\epsilon_u}\right)\left[\frac{256}{3}\frac{1}{\epsilon_v^4}+\frac{64}{5}\frac{\hat{n}_1^2}{\hat{m}_1^4}-\frac{64}{5}\frac{\hat{n}_2}{\hat{m}_1^3}+\mathcal{O}(\epsilon_v^2)\right]+\mathcal{O}(\epsilon_u^2).
\end{align}
It is then straightforward to see that $S_{\rm EE}$ is given by \eq{RT-Area}.

%%%%%%%%%%%%%%%%%%%%%%%%%%%%%%%%%%%%%%%%%%%%%%%%%%
%%%%%%%%%%%%%%%%%%%%%%%%%%%%%%%%%%%%%%%%%%%%%%%%%%
\bibliographystyle{JHEP}
\bibliography{large2small}

\providecommand{\href}[2]{#2}\begingroup\raggedright\begin{thebibliography}{10}

\bibitem{Nahm:1977tg}
W.~Nahm, \emph{{Supersymmetries and their Representations}},
  \href{https://doi.org/10.1016/0550-3213(78)90218-3}{\emph{Nucl. Phys. B}
  {\bfseries 135} (1978) 149}.

\bibitem{Antoniadis:1998ki}
I.~Antoniadis, E.~Dudas and A.~Sagnotti, \emph{{Supersymmetry breaking, open
  strings and M theory}},
  \href{https://doi.org/10.1016/S0550-3213(98)00806-2}{\emph{Nucl. Phys. B}
  {\bfseries 544} (1999) 469}
  [\href{https://arxiv.org/abs/hep-th/9807011}{{\ttfamily hep-th/9807011}}].

\bibitem{Antoniadis:1998ep}
I.~Antoniadis, G.~D'Appollonio, E.~Dudas and A.~Sagnotti, \emph{{Partial
  breaking of supersymmetry, open strings and M theory}},
  \href{https://doi.org/10.1016/S0550-3213(99)00232-1}{\emph{Nucl. Phys. B}
  {\bfseries 553} (1999) 133}
  [\href{https://arxiv.org/abs/hep-th/9812118}{{\ttfamily hep-th/9812118}}].

\bibitem{Cordova:2016xhm}
C.~Cordova, T.~T. Dumitrescu and K.~Intriligator, \emph{{Deformations of
  Superconformal Theories}},
  \href{https://doi.org/10.1007/JHEP11(2016)135}{\emph{JHEP} {\bfseries 11}
  (2016) 135} [\href{https://arxiv.org/abs/1602.01217}{{\ttfamily
  1602.01217}}].

\bibitem{Blum:1997fw}
J.~D. Blum and K.~A. Intriligator, \emph{{Consistency conditions for branes at
  orbifold singularities}},
  \href{https://doi.org/10.1016/S0550-3213(97)00450-1}{\emph{Nucl. Phys. B}
  {\bfseries 506} (1997) 223}
  [\href{https://arxiv.org/abs/hep-th/9705030}{{\ttfamily hep-th/9705030}}].

\bibitem{Blum:1997mm}
J.~D. Blum and K.~A. Intriligator, \emph{{New phases of string theory and 6-D
  RG fixed points via branes at orbifold singularities}},
  \href{https://doi.org/10.1016/S0550-3213(97)00449-5}{\emph{Nucl. Phys. B}
  {\bfseries 506} (1997) 199}
  [\href{https://arxiv.org/abs/hep-th/9705044}{{\ttfamily hep-th/9705044}}].

\bibitem{Brunner:1997gf}
I.~Brunner and A.~Karch, \emph{{Branes at orbifolds versus Hanany Witten in
  six-dimensions}},
  \href{https://doi.org/10.1088/1126-6708/1998/03/003}{\emph{JHEP} {\bfseries
  03} (1998) 003} [\href{https://arxiv.org/abs/hep-th/9712143}{{\ttfamily
  hep-th/9712143}}].

\bibitem{Intriligator:1997dh}
K.~A. Intriligator, \emph{{New string theories in six-dimensions via branes at
  orbifold singularities}},
  \href{https://doi.org/10.4310/ATMP.1997.v1.n2.a5}{\emph{Adv. Theor. Math.
  Phys.} {\bfseries 1} (1998) 271}
  [\href{https://arxiv.org/abs/hep-th/9708117}{{\ttfamily hep-th/9708117}}].

\bibitem{Gaiotto:2009we}
D.~Gaiotto, \emph{{N=2 dualities}},
  \href{https://doi.org/10.1007/JHEP08(2012)034}{\emph{JHEP} {\bfseries 08}
  (2012) 034} [\href{https://arxiv.org/abs/0904.2715}{{\ttfamily 0904.2715}}].

\bibitem{Dimofte:2011ju}
T.~Dimofte, D.~Gaiotto and S.~Gukov, \emph{{Gauge Theories Labelled by
  Three-Manifolds}},
  \href{https://doi.org/10.1007/s00220-013-1863-2}{\emph{Commun. Math. Phys.}
  {\bfseries 325} (2014) 367}
  [\href{https://arxiv.org/abs/1108.4389}{{\ttfamily 1108.4389}}].

\bibitem{Duff:1990xz}
M.~J. Duff and K.~S. Stelle, \emph{{Multimembrane solutions of D = 11
  supergravity}},
  \href{https://doi.org/10.1016/0370-2693(91)91371-2}{\emph{Phys. Lett. B}
  {\bfseries 253} (1991) 113}.

\bibitem{Gueven:1992hh}
R.~Gueven, \emph{{Black p-brane solutions of D = 11 supergravity theory}},
  \href{https://doi.org/10.1016/0370-2693(92)90540-K}{\emph{Phys. Lett. B}
  {\bfseries 276} (1992) 49}.

\bibitem{Strominger:1995ac}
A.~Strominger, \emph{{Open p-branes}},
  \href{https://doi.org/10.1016/0370-2693(96)00712-5}{\emph{Phys. Lett. B}
  {\bfseries 383} (1996) 44}
  [\href{https://arxiv.org/abs/hep-th/9512059}{{\ttfamily hep-th/9512059}}].

\bibitem{Witten:1996hc}
E.~Witten, \emph{{Five-brane effective action in M theory}},
  \href{https://doi.org/10.1016/S0393-0440(97)80160-X}{\emph{J. Geom. Phys.}
  {\bfseries 22} (1997) 103}
  [\href{https://arxiv.org/abs/hep-th/9610234}{{\ttfamily hep-th/9610234}}].

\bibitem{Maldacena:1997re}
J.~M. Maldacena, \emph{{The Large N limit of superconformal field theories and
  supergravity}}, \href{https://doi.org/10.1023/A:1026654312961,
  10.4310/ATMP.1998.v2.n2.a1}{\emph{Int. J. Theor. Phys.} {\bfseries 38} (1999)
  1113} [\href{https://arxiv.org/abs/hep-th/9711200}{{\ttfamily
  hep-th/9711200}}].

\bibitem{Witten:1998wy}
E.~Witten, \emph{{AdS / CFT correspondence and topological field theory}},
  \href{https://doi.org/10.1088/1126-6708/1998/12/012}{\emph{JHEP} {\bfseries
  12} (1998) 012} [\href{https://arxiv.org/abs/hep-th/9812012}{{\ttfamily
  hep-th/9812012}}].

\bibitem{Bullimore:2014upa}
M.~Bullimore and H.-C. Kim, \emph{{The Superconformal Index of the (2,0) Theory
  with Defects}}, \href{https://doi.org/10.1007/JHEP05(2015)048}{\emph{JHEP}
  {\bfseries 05} (2015) 048} [\href{https://arxiv.org/abs/1412.3872}{{\ttfamily
  1412.3872}}].

\bibitem{Beem:2015aoa}
C.~Beem, M.~Lemos, L.~Rastelli and B.~C. van Rees, \emph{{The (2, 0)
  superconformal bootstrap}},
  \href{https://doi.org/10.1103/PhysRevD.93.025016}{\emph{Phys. Rev. D}
  {\bfseries 93} (2016) 025016}
  [\href{https://arxiv.org/abs/1507.05637}{{\ttfamily 1507.05637}}].

\bibitem{Heckman:2015bfa}
J.~J. Heckman, D.~R. Morrison, T.~Rudelius and C.~Vafa, \emph{{Atomic
  Classification of 6D SCFTs}},
  \href{https://doi.org/10.1002/prop.201500024}{\emph{Fortsch. Phys.}
  {\bfseries 63} (2015) 468}
  [\href{https://arxiv.org/abs/1502.05405}{{\ttfamily 1502.05405}}].

\bibitem{Apruzzi:2013yva}
F.~Apruzzi, M.~Fazzi, D.~Rosa and A.~Tomasiello, \emph{{All AdS$_7$ solutions
  of type II supergravity}},
  \href{https://doi.org/10.1007/JHEP04(2014)064}{\emph{JHEP} {\bfseries 04}
  (2014) 064} [\href{https://arxiv.org/abs/1309.2949}{{\ttfamily 1309.2949}}].

\bibitem{Henningson:1998gx}
M.~Henningson and K.~Skenderis, \emph{{The Holographic Weyl anomaly}},
  \href{https://doi.org/10.1088/1126-6708/1998/07/023}{\emph{JHEP} {\bfseries
  07} (1998) 023} [\href{https://arxiv.org/abs/hep-th/9806087}{{\ttfamily
  hep-th/9806087}}].

\bibitem{Hung:2011xb}
L.-Y. Hung, R.~C. Myers and M.~Smolkin, \emph{{On Holographic Entanglement
  Entropy and Higher Curvature Gravity}},
  \href{https://doi.org/10.1007/JHEP04(2011)025}{\emph{JHEP} {\bfseries 04}
  (2011) 025} [\href{https://arxiv.org/abs/1101.5813}{{\ttfamily 1101.5813}}].

\bibitem{Andrei:2018die}
N.~Andrei et~al., \emph{{Boundary and Defect CFT: Open Problems and
  Applications}}, \href{https://doi.org/10.1088/1751-8121/abb0fe}{\emph{J.
  Phys. A} {\bfseries 53} (2020) 453002}
  [\href{https://arxiv.org/abs/1810.05697}{{\ttfamily 1810.05697}}].

\bibitem{Soderberg:2017oaa}
A.~S\"oderberg, \emph{{Anomalous Dimensions in the WF O($N$) Model with a
  Monodromy Line Defect}},
  \href{https://doi.org/10.1007/JHEP03(2018)058}{\emph{JHEP} {\bfseries 03}
  (2018) 058} [\href{https://arxiv.org/abs/1706.02414}{{\ttfamily
  1706.02414}}].

\bibitem{Lauria:2020emq}
E.~Lauria, P.~Liendo, B.~C. Van~Rees and X.~Zhao, \emph{{Line and surface
  defects for the free scalar field}},
  \href{https://doi.org/10.1007/JHEP01(2021)060}{\emph{JHEP} {\bfseries 01}
  (2021) 060} [\href{https://arxiv.org/abs/2005.02413}{{\ttfamily
  2005.02413}}].

\bibitem{Giombi:2021uae}
S.~Giombi, E.~Helfenberger, Z.~Ji and H.~Khanchandani, \emph{{Monodromy defects
  from hyperbolic space}},
  \href{https://doi.org/10.1007/JHEP02(2022)041}{\emph{JHEP} {\bfseries 02}
  (2022) 041} [\href{https://arxiv.org/abs/2102.11815}{{\ttfamily
  2102.11815}}].

\bibitem{Bianchi:2021snj}
L.~Bianchi, A.~Chalabi, V.~Proch\'azka, B.~Robinson and J.~Sisti,
  \emph{{Monodromy defects in free field theories}},
  \href{https://doi.org/10.1007/JHEP08(2021)013}{\emph{JHEP} {\bfseries 08}
  (2021) 013} [\href{https://arxiv.org/abs/2104.01220}{{\ttfamily
  2104.01220}}].

\bibitem{Gukov:2006jk}
S.~Gukov and E.~Witten, \emph{{Gauge Theory, Ramification, And The Geometric
  Langlands Program}},  \href{https://arxiv.org/abs/hep-th/0612073}{{\ttfamily
  hep-th/0612073}}.

\bibitem{Gukov:2008sn}
S.~Gukov and E.~Witten, \emph{{Rigid Surface Operators}},
  \href{https://doi.org/10.4310/ATMP.2010.v14.n1.a3}{\emph{Adv. Theor. Math.
  Phys.} {\bfseries 14} (2010) 87}
  [\href{https://arxiv.org/abs/0804.1561}{{\ttfamily 0804.1561}}].

\bibitem{Alday:2009fs}
L.~F. Alday, D.~Gaiotto, S.~Gukov, Y.~Tachikawa and H.~Verlinde, \emph{{Loop
  and surface operators in N=2 gauge theory and Liouville modular geometry}},
  \href{https://doi.org/10.1007/JHEP01(2010)113}{\emph{JHEP} {\bfseries 01}
  (2010) 113} [\href{https://arxiv.org/abs/0909.0945}{{\ttfamily 0909.0945}}].

\bibitem{Ganor:1996nf}
O.~J. Ganor, \emph{{Six-dimensional tensionless strings in the large N limit}},
  \href{https://doi.org/10.1016/S0550-3213(96)00702-X}{\emph{Nucl. Phys. B}
  {\bfseries 489} (1997) 95}
  [\href{https://arxiv.org/abs/hep-th/9605201}{{\ttfamily hep-th/9605201}}].

\bibitem{Hung:2014npa}
L.-Y. Hung, R.~C. Myers and M.~Smolkin, \emph{{Twist operators in higher
  dimensions}}, \href{https://doi.org/10.1007/JHEP10(2014)178}{\emph{JHEP}
  {\bfseries 10} (2014) 178} [\href{https://arxiv.org/abs/1407.6429}{{\ttfamily
  1407.6429}}].

\bibitem{Lewkowycz:2014jia}
A.~Lewkowycz and E.~Perlmutter, \emph{{Universality in the geometric dependence
  of Renyi entropy}},
  \href{https://doi.org/10.1007/JHEP01(2015)080}{\emph{JHEP} {\bfseries 01}
  (2015) 080} [\href{https://arxiv.org/abs/1407.8171}{{\ttfamily 1407.8171}}].

\bibitem{Billo:2016cpy}
M.~Bill\`o, V.~Gon\c{c}alves, E.~Lauria and M.~Meineri, \emph{{Defects in
  conformal field theory}},
  \href{https://doi.org/10.1007/JHEP04(2016)091}{\emph{JHEP} {\bfseries 04}
  (2016) 091} [\href{https://arxiv.org/abs/1601.02883}{{\ttfamily
  1601.02883}}].

\bibitem{Jensen:2015swa}
K.~Jensen and A.~O'Bannon, \emph{{Constraint on Defect and Boundary
  Renormalization Group Flows}},
  \href{https://doi.org/10.1103/PhysRevLett.116.091601}{\emph{Phys. Rev. Lett.}
  {\bfseries 116} (2016) 091601}
  [\href{https://arxiv.org/abs/1509.02160}{{\ttfamily 1509.02160}}].

\bibitem{Kobayashi:2018lil}
N.~Kobayashi, T.~Nishioka, Y.~Sato and K.~Watanabe, \emph{{Towards a
  $C$-theorem in defect CFT}},
  \href{https://doi.org/10.1007/JHEP01(2019)039}{\emph{JHEP} {\bfseries 01}
  (2019) 039} [\href{https://arxiv.org/abs/1810.06995}{{\ttfamily
  1810.06995}}].

\bibitem{Bachas:2013vza}
C.~Bachas, E.~D'Hoker, J.~Estes and D.~Krym, \emph{{M-theory Solutions
  Invariant under $D(2,1;\gamma) \oplus D(2,1;\gamma)$}},
  \href{https://doi.org/10.1002/prop.201300039}{\emph{Fortsch. Phys.}
  {\bfseries 62} (2014) 207} [\href{https://arxiv.org/abs/1312.5477}{{\ttfamily
  1312.5477}}].

\bibitem{Bobev:2013yra}
N.~Bobev, K.~Pilch and N.~P. Warner, \emph{{Supersymmetric Janus Solutions in
  Four Dimensions}}, \href{https://doi.org/10.1007/JHEP06(2014)058}{\emph{JHEP}
  {\bfseries 06} (2014) 058} [\href{https://arxiv.org/abs/1311.4883}{{\ttfamily
  1311.4883}}].

\bibitem{DHoker:2009lky}
E.~D'Hoker, J.~Estes, M.~Gutperle and D.~Krym, \emph{{Janus solutions in
  M-theory}}, \href{https://doi.org/10.1088/1126-6708/2009/06/018}{\emph{JHEP}
  {\bfseries 06} (2009) 018} [\href{https://arxiv.org/abs/0904.3313}{{\ttfamily
  0904.3313}}].

\bibitem{Howe:1997ue}
P.~S. Howe, N.~D. Lambert and P.~C. West, \emph{{The Selfdual string soliton}},
  \href{https://doi.org/10.1016/S0550-3213(97)00750-5}{\emph{Nucl. Phys. B}
  {\bfseries 515} (1998) 203}
  [\href{https://arxiv.org/abs/hep-th/9709014}{{\ttfamily hep-th/9709014}}].

\bibitem{Maldacena:1998im}
J.~M. Maldacena, \emph{{Wilson loops in large N field theories}},
  \href{https://doi.org/10.1103/PhysRevLett.80.4859}{\emph{Phys. Rev. Lett.}
  {\bfseries 80} (1998) 4859}
  [\href{https://arxiv.org/abs/hep-th/9803002}{{\ttfamily hep-th/9803002}}].

\bibitem{Berenstein:1998ij}
D.~E. Berenstein, R.~Corrado, W.~Fischler and J.~M. Maldacena, \emph{{The
  Operator product expansion for Wilson loops and surfaces in the large N
  limit}}, \href{https://doi.org/10.1103/PhysRevD.59.105023}{\emph{Phys. Rev.
  D} {\bfseries 59} (1999) 105023}
  [\href{https://arxiv.org/abs/hep-th/9809188}{{\ttfamily hep-th/9809188}}].

\bibitem{Gentle:2015jma}
S.~A. Gentle, M.~Gutperle and C.~Marasinou, \emph{{Entanglement entropy of
  Wilson surfaces from bubbling geometries in M-theory}},
  \href{https://doi.org/10.1007/JHEP08(2015)019}{\emph{JHEP} {\bfseries 08}
  (2015) 019} [\href{https://arxiv.org/abs/1506.00052}{{\ttfamily
  1506.00052}}].

\bibitem{Estes:2018tnu}
J.~Estes, D.~Krym, A.~O'Bannon, B.~Robinson and R.~Rodgers, \emph{{Wilson
  Surface Central Charge from Holographic Entanglement Entropy}},
  \href{https://doi.org/10.1007/JHEP05(2019)032}{\emph{JHEP} {\bfseries 05}
  (2019) 032} [\href{https://arxiv.org/abs/1812.00923}{{\ttfamily
  1812.00923}}].

\bibitem{Jensen:2018rxu}
K.~Jensen, A.~O'Bannon, B.~Robinson and R.~Rodgers, \emph{{From the Weyl
  Anomaly to Entropy of Two-Dimensional Boundaries and Defects}},
  \href{https://doi.org/10.1103/PhysRevLett.122.241602}{\emph{Phys. Rev. Lett.}
  {\bfseries 122} (2019) 241602}
  [\href{https://arxiv.org/abs/1812.08745}{{\ttfamily 1812.08745}}].

\bibitem{Chalabi:2020iie}
A.~Chalabi, A.~O'Bannon, B.~Robinson and J.~Sisti, \emph{{Central charges of 2d
  superconformal defects}},
  \href{https://doi.org/10.1007/JHEP05(2020)095}{\emph{JHEP} {\bfseries 05}
  (2020) 095} [\href{https://arxiv.org/abs/2003.02857}{{\ttfamily
  2003.02857}}].

\bibitem{Drukker:2020atp}
N.~Drukker, M.~Probst and M.~Tr\'epanier, \emph{{Defect CFT techniques in the
  6d $\mathcal{N} = (2,0)$ theory}},
  \href{https://doi.org/10.1007/JHEP03(2021)261}{\emph{JHEP} {\bfseries 03}
  (2021) 261} [\href{https://arxiv.org/abs/2009.10732}{{\ttfamily
  2009.10732}}].

\bibitem{Drukker:2020dcz}
N.~Drukker, M.~Probst and M.~Tr\'epanier, \emph{{Surface operators in the 6d N
  = (2, 0) theory}}, \href{https://doi.org/10.1088/1751-8121/aba1b7}{\emph{J.
  Phys. A} {\bfseries 53} (2020) 365401}
  [\href{https://arxiv.org/abs/2003.12372}{{\ttfamily 2003.12372}}].

\bibitem{Faedo:2020nol}
F.~Faedo, Y.~Lozano and N.~Petri, \emph{{Searching for surface defect CFTs
  within AdS$_3$}}, \href{https://doi.org/10.1007/JHEP11(2020)052}{\emph{JHEP}
  {\bfseries 11} (2020) 052}
  [\href{https://arxiv.org/abs/2007.16167}{{\ttfamily 2007.16167}}].

\bibitem{frappat1996dictionary}
L.~Frappat, A.~Sciarrino and P.~Sorba, \emph{{Dictionary on Lie
  Superalgebras}},  1996.

\bibitem{D_Hoker_2008}
E.~D’Hoker, J.~Estes, M.~Gutperle, D.~Krym and P.~Sorba, \emph{Half-bps
  supergravity solutions and superalgebras},
  \href{https://doi.org/10.1088/1126-6708/2008/12/047}{\emph{Journal of High
  Energy Physics} {\bfseries 2008} (2008) 047–047}.

\bibitem{D_Hoker_2008b}
E.~D{\textquotesingle}Hoker, J.~Estes, M.~Gutperle and D.~Krym, \emph{{Exact
  half-{BPS} flux solutions in M-theory I: local solutions}},
  \href{https://doi.org/10.1088/1126-6708/2008/08/028}{\emph{Journal of High
  Energy Physics} {\bfseries 2008} (2008) 028}.

\bibitem{Dibitetto:2017tve}
G.~Dibitetto and N.~Petri, \emph{{BPS objects in D = 7 supergravity and their
  M-theory origin}}, \href{https://doi.org/10.1007/JHEP12(2017)041}{\emph{JHEP}
  {\bfseries 12} (2017) 041}
  [\href{https://arxiv.org/abs/1707.06152}{{\ttfamily 1707.06152}}].

\bibitem{Dibitetto:2017klx}
G.~Dibitetto and N.~Petri, \emph{{6d surface defects from massive type IIA}},
  \href{https://doi.org/10.1007/JHEP01(2018)039}{\emph{JHEP} {\bfseries 01}
  (2018) 039} [\href{https://arxiv.org/abs/1707.06154}{{\ttfamily
  1707.06154}}].

\bibitem{Izquierdo:1995ms}
J.~M. Izquierdo, N.~D. Lambert, G.~Papadopoulos and P.~K. Townsend,
  \emph{{Dyonic membranes}},
  \href{https://doi.org/10.1016/0550-3213(95)00606-0}{\emph{Nucl. Phys. B}
  {\bfseries 460} (1996) 560}
  [\href{https://arxiv.org/abs/hep-th/9508177}{{\ttfamily hep-th/9508177}}].

\bibitem{Lozano:2020bxo}
Y.~Lozano, C.~Nunez, A.~Ramirez and S.~Speziali, \emph{{$M$-strings and AdS$_3$
  solutions to M-theory with small $\mathcal{N}=(0,4)$ supersymmetry}},
  \href{https://doi.org/10.1007/JHEP08(2020)118}{\emph{JHEP} {\bfseries 08}
  (2020) 118} [\href{https://arxiv.org/abs/2005.06561}{{\ttfamily
  2005.06561}}].

\bibitem{Estes:2012vm}
J.~Estes, R.~Feldman and D.~Krym, \emph{{Exact half-BPS flux solutions in $M$
  theory with D(2,1;$c^\prime$;0)$^2$ symmetry: Local solutions}},
  \href{https://doi.org/10.1103/PhysRevD.87.046008}{\emph{Phys. Rev. D}
  {\bfseries 87} (2013) 046008}
  [\href{https://arxiv.org/abs/1209.1845}{{\ttfamily 1209.1845}}].

\bibitem{DHoker:2008rje}
E.~D'Hoker, J.~Estes, M.~Gutperle and D.~Krym, \emph{{Exact Half-BPS Flux
  Solutions in M-theory II: Global solutions asymptotic to AdS$_7\times$
  S$^4$}}, \href{https://doi.org/10.1088/1126-6708/2008/12/044}{\emph{JHEP}
  {\bfseries 12} (2008) 044} [\href{https://arxiv.org/abs/0810.4647}{{\ttfamily
  0810.4647}}].

\bibitem{Ryu:2006bv}
S.~Ryu and T.~Takayanagi, \emph{{Holographic derivation of entanglement entropy
  from AdS/CFT}},
  \href{https://doi.org/10.1103/PhysRevLett.96.181602}{\emph{Phys. Rev. Lett.}
  {\bfseries 96} (2006) 181602}
  [\href{https://arxiv.org/abs/hep-th/0603001}{{\ttfamily hep-th/0603001}}].

\bibitem{Ryu:2006ef}
S.~Ryu and T.~Takayanagi, \emph{{Aspects of Holographic Entanglement Entropy}},
  \href{https://doi.org/10.1088/1126-6708/2006/08/045}{\emph{JHEP} {\bfseries
  08} (2006) 045} [\href{https://arxiv.org/abs/hep-th/0605073}{{\ttfamily
  hep-th/0605073}}].

\bibitem{Nishioka:2021uef}
T.~Nishioka and Y.~Sato, \emph{{Free energy and defect $C$-theorem in free
  scalar theory}}, \href{https://doi.org/10.1007/JHEP05(2021)074}{\emph{JHEP}
  {\bfseries 05} (2021) 074}
  [\href{https://arxiv.org/abs/2101.02399}{{\ttfamily 2101.02399}}].

\bibitem{Holzhey:1994we}
C.~Holzhey, F.~Larsen and F.~Wilczek, \emph{{Geometric and renormalized entropy
  in conformal field theory}},
  \href{https://doi.org/10.1016/0550-3213(94)90402-2}{\emph{Nucl. Phys. B}
  {\bfseries 424} (1994) 443}
  [\href{https://arxiv.org/abs/hep-th/9403108}{{\ttfamily hep-th/9403108}}].

\bibitem{Estes:2014hka}
J.~Estes, K.~Jensen, A.~O'Bannon, E.~Tsatis and T.~Wrase, \emph{{On Holographic
  Defect Entropy}}, \href{https://doi.org/10.1007/JHEP05(2014)084}{\emph{JHEP}
  {\bfseries 05} (2014) 084} [\href{https://arxiv.org/abs/1403.6475}{{\ttfamily
  1403.6475}}].

\bibitem{Jensen:2013lxa}
K.~Jensen and A.~O'Bannon, \emph{{Holography, Entanglement Entropy, and
  Conformal Field Theories with Boundaries or Defects}},
  \href{https://doi.org/10.1103/PhysRevD.88.106006}{\emph{Phys. Rev. D}
  {\bfseries 88} (2013) 106006}
  [\href{https://arxiv.org/abs/1309.4523}{{\ttfamily 1309.4523}}].

\bibitem{deHaro:2000vlm}
S.~de~Haro, S.~N. Solodukhin and K.~Skenderis, \emph{{Holographic
  reconstruction of space-time and renormalization in the AdS / CFT
  correspondence}}, \href{https://doi.org/10.1007/s002200100381}{\emph{Commun.
  Math. Phys.} {\bfseries 217} (2001) 595}
  [\href{https://arxiv.org/abs/hep-th/0002230}{{\ttfamily hep-th/0002230}}].

\bibitem{Skenderis:2006uy}
K.~Skenderis and M.~Taylor, \emph{{Kaluza-Klein holography}},
  \href{https://doi.org/10.1088/1126-6708/2006/05/057}{\emph{JHEP} {\bfseries
  05} (2006) 057} [\href{https://arxiv.org/abs/hep-th/0603016}{{\ttfamily
  hep-th/0603016}}].

\bibitem{Bianchi:2019sxz}
L.~Bianchi and M.~Lemos, \emph{{Superconformal surfaces in four dimensions}},
  \href{https://doi.org/10.1007/JHEP06(2020)056}{\emph{JHEP} {\bfseries 06}
  (2020) 056} [\href{https://arxiv.org/abs/1911.05082}{{\ttfamily
  1911.05082}}].

\bibitem{Capuozzo:2023fll}
P.~Capuozzo, J.~Estes, B.~Robinson and B.~Suzzoni, \emph{{Holographic Weyl
  Anomalies for 4d Defects in 6d SCFTs}},
  \href{https://arxiv.org/abs/2310.17447}{{\ttfamily 2310.17447}}.

\bibitem{Meneghelli:2022gps}
C.~Meneghelli and M.~Tr\'epanier, \emph{{Bootstrapping string dynamics in the
  6d \ensuremath{\mathcal{N}} = (2, 0) theories}},
  \href{https://doi.org/10.1007/JHEP07(2023)165}{\emph{JHEP} {\bfseries 07}
  (2023) 165} [\href{https://arxiv.org/abs/2212.05020}{{\ttfamily
  2212.05020}}].

\end{thebibliography}\endgroup
%%%%%%%%%%%%%%%%%%%%%%%%%%%%%%%%%%%%%%%%%%%%%%%%%%
%%%%%%%%%%%%%%%%%%%%%%%%%%%%%%%%%%%%%%%%%%%%%%%%%%

\end{document}